\documentclass[twocolumn, tighten, twocolappendix]{aastex631}

\newcommand{\sersic}{S$\rm{\acute{e}}$rsic}

\usepackage{amsmath}

\newcommand{\dnbreak}{$\rm {D}_n(4000)$}

\newcommand{\ha}{$\rm H{\alpha}$}
\newcommand{\hb}{$\rm H{\beta}$}
\newcommand{\hd}{$\rm H{\delta}_{A}$}
\newcommand{\haew}{$\rm {EW}({H\alpha})$}

\newcommand{\hdew}{$\rm {EW}({H\delta}_{A})$}
\newcommand{\hi}{H{\sc i}}

\newcommand{\ebv}{$E(B-V)$}

\newcommand{\hasb}{$\Sigma_{\rm H\alpha}$}

\newcommand{\f}{$f$}

\newcommand{\oiii}{[O{\sc iii}]}

\newcommand{\nii}{[N{\sc ii}]}

\newcommand{\vstar}{$v_\ast$}
\newcommand{\sigstar}{$\sigma_\ast$}
\newcommand{\Sigsfr}{$\rm \Sigma_{SFR}$}
\newcommand{\Sigha}{$\rm \Sigma_{H\alpha}$}
\newcommand{\mstar}{$M_\ast$}
\newcommand{\Sigstar}{$\Sigma_\ast$}
\newcommand{\Sigmol}{$\rm \Sigma_{H2}$}
\newcommand{\Siggas}{$\rm \Sigma_{gas}$}
\newcommand{\SigHI}{$\rm \Sigma_{HI}$}

\newcommand{\Qtracer}{\dnbreak\ - $\log$ \haew}
\newcommand{\fq}{$f_Q$}
\defcitealias{2018ApJ...856..137W}{Paper I}

\shorttitle{The origin of quenched but gas-rich regions}
\shortauthors{Jing et al.}
\graphicspath{{./}{figures/}}

\begin{document}

\title{On the Origin of Quenched but Gas-rich Regions at Kiloparsec Scales in Nearby Galaxies}

\correspondingauthor{Tao Jing}
\email{jingt20@mails.tsinghua.edu.cn}
\correspondingauthor{Cheng Li}
\email{cli2015@tsinghua.edu.cn}

\author[0009-0004-6271-4321]{Tao Jing}
\affiliation{Department of Astronomy, Tsinghua University, Beijing 100084, China}

\author[0000-0002-8711-8970]{Cheng Li}
\affiliation{Department of Astronomy, Tsinghua University, Beijing 100084, China}

%\author{Niu Li}
%\affiliation{Department of Astronomy, Tsinghua University, Beijing 100084, China}

%\author{Others}

\begin{abstract}
We use resolved spectroscopy from MaNGA to investigate the significance 
of both local and global properties of galaxies to the cessation of star
formation at kpc scales. Quenched regions are identified from a sample of 
isolated disk galaxies by a single-parameter criterion 
\Qtracer$~>1.6-\log 2=1.3$, and are divided into gas-rich quenched 
regions (GRQRs) and gas-poor quenched regions (GPQRs) according to 
the surface density of cold gas (\Siggas). Both types of quenched regions 
tend to be hosted by non-AGN galaxies with relatively high mass 
(\mstar$\ga 10^{10}M_\odot$) and red colors (${\rm NUV}-r\ga 3$), 
as well as low star formation rate and high central density at fixed mass. 
They span wide ranges in other properties including structural parameters 
that are similar to the parent sample, indicating that the conditions responsible 
for quenching in gas-rich regions are largely independent on the global 
properties of galaxies. We train random forest (RF) classifiers and regressors
for predicting quenching in our sample with 15 local/global properties. 
\Sigstar\ is the most important property for quenching, especially for GRQRs.
These results strongly indicate the important roles of low-mass hot evolved 
stars which are numerous and long-lived in quenched regions and can provide 
substantial radiation pressure to support the surrounding gas against gravitational
collapse.  The different feature importance for quenching as found previously 
by \citet{Bluck2020-global-local,Bluck2020-central-satellite} are partly due to the 
different definitions of quenched regions, particularly the different requirements 
on \haew. 
% As the quenched region selected with \haew\ requirements is a subsample of that without \haew\ requirements, it suggests AGN feedback cannot be deemed the sole quenching mechanism.
\end{abstract}


\section{Introduction} \label{sec:intro}

%Based on the activity of star formation, the galaxies can be separated into two groups: star-forming and quiescence galaxies. The star formation situation is widely related to other galaxy properties, of which the galaxies have a significant bimodal distribution. A star-forming galaxy usually has bluer color, less mass, more gas, the Sd/c/b/a morphology according to Hubble classification, and tends to locate in a denser environment. In addition to differences in the formation of primordial galaxies, the existence of such two groups of galaxies are also due to galaxy evolution, which means that some of the quiescence galaxies are transformed from star-forming phase by some rapid quenching processes. But the details of these processes are unclear up to now. 

Observations of galaxy populations at different redshifts have well 
established that the fraction of red quiescent galaxies has significantly 
increased since a redshift of unity 
\citep[e.g.][]{2004ApJ...600L..11B,2006ApJ...651..120B,2007ApJ...665..265F},
implying that the cessation of star formation has been a driving process 
in galaxy evolution over the past $\sim8$ Gyr.  However, how galaxies 
quench their star formation remains an unresolved problem, despite decades 
of studies which have proposed a variety of quenching mechanisms. 
In most cases, star formation ceases in a galaxy through removal or 
heating of cold gas as a result of processes external or internal to the galaxy. 
External processes include shock heating of gas in massive dark matter halos 
\citep[e.g.][]{1977MNRAS.179..541R,1977ApJ...211..638S,1984Natur.311..517B,2003MNRAS.345..349B,2005MNRAS.363....2K,2006MNRAS.370.1651C,2006MNRAS.368....2D,2007MNRAS.380..339B,2008MNRAS.389..567C,2008MNRAS.390.1326O},
stripping of hot and even cold gas by ram pressure, tidal interactions, and 
strangulation
\citep[e.g.][]{1972ApJ...176....1G,1972ApJ...178..623T,1996Natur.379..613M,
	1999MNRAS.308..947A,2000ApJ...540..113B,2004AJ....127.3300V,
	2009AJ....138.1741C,2009MNRAS.394.1213W,Li2012,
	2013MNRAS.429.1747M,Zhang2013,2014MNRAS.438..444B,
	2015MNRAS.449.3503D,2015Natur.521..192P,2019MNRAS.483.5444D,
	2017ApJ...844...48P,2021ApJ...919..134S}, and major mergers of gas-rich galaxies 
which induce fast conversion of cold gas to stars at galactic center as well as 
the so-called ``quasar-mode'' active galactic nucleus (AGN) feedback 
\citep[e.g.][]{1991ApJ...370L..65B,1996ApJ...464..641M,2005Natur.433..604D,2005ApJ...620L..79S,2006ApJS..163....1H,2009ApJ...691.1168H,2012ARA&A..50..455F}. 
Internal to a galaxy, the processes proposed for quenching include stellar 
feedback driven by radiation pressure, stellar wind and supernova explosions
\cite[e.g.][]{1974ApJ...189L.105C,1977ApJ...218..148M,1979ARA&A..17..213M,1979MNRAS.186...59W,1986ApJ...303...39D,Binette1994,2009ApJ...695..292C,2015ApJ...803...77C,2015MNRAS.454.2691M,2016ApJ...829..130R,2018MNRAS.481.1774H,2018MNRAS.477.1578H,2018MNRAS.480..800H,2020ApJ...898...23L,2022ApJ...936..137O}, the ``radio-mode'' AGN feedback 
from accretion of hot halo gas onto the central supermassive black hole
in massive central galaxies \citep[e.g.][]{2006MNRAS.370..645B,2006MNRAS.365...11C,2008MNRAS.390.1399B,2012ARA&A..50..455F}, and 
``compaction quenching'' which explains the formation of compact quenched 
red nuggets at high redshift by accretion-driven violent instability of gas-rich discs
\citep[e.g.][]{2014MNRAS.438.1870D,2015MNRAS.450.2327Z}.

Observations have also clearly established that the majority of quenched 
galaxies at the center of galaxy groups/clusters and at both low and high 
redshifts have a massive dense structure at the galactic center such as 
a prominent bulge
\citep[e.g.][]{2008ApJ...682..355B,2010ApJ...719.1969B,2010MNRAS.405..783M,2012ApJ...760..131C,2013ApJ...776...63F,2014MNRAS.441..599B,2017ApJ...840...47B}.
In addition, a significant population of red-sequence galaxies 
at both low and high redshifts have been observed to present disk-dominated spirals 
with large bulges in the center 
\citep[e.g.][]{2005A&A...443..435W,2009MNRAS.393.1302W,2009MNRAS.393.1324B,2010ApJ...719.1969B,2010MNRAS.405..783M,Hao2019,Guo2020,Cui2024}. 
These findings support the scenario of ``morphological quenching'' as originally 
proposed by \cite{2009ApJ...707..250M}. In this picture, a massive bulge, 
or more generally a centrally-concentrated mass distribution may effectively 
stabilize the gas disk, thus suppressing or even quenching the star formation
by sheared perturbations within the disk. As a natural prediction of this process,
the galaxy may sustain a large amount of cold gas in the disk even when the star 
formation is totally quenched. Indeed, high detection rates of \hi\ emission 
have been reported in massive red spirals \citep[e.g.][]{Guo2020,Wang2022-FAST}
and early-type galaxies \citep[e.g.][]{2012MNRAS.422.1835S}. 
As pointed out by \cite{2021ApJ...916...38Z},
however, many of the massive red spirals as selected by the optical 
color index $u-r$ in previous studies are actually green or even blue in the 
NUV-to-optical color ${\rm NUV}-r$. In a recent study, \cite{Li2023} identified 
a sample of 47 quenched \hi-rich galaxies that are truly red with ${\rm NUV}-r>5$ 
and have unusually large amounts of total \hi\ mass. The authors found no evidence 
for morphological quenching, however, by comparing the structural properties 
of the red \hi-rich galaxies with control samples of \hi-normal galaxies.
In addition, by comparing the red \hi-rich galaxies with a complete sample 
selected from the Sloan Digital Sky Survey \citep{2000AJ....120.1579Y} as 
well as previous samples of \hi-rich galaxies with low or suppressed 
star formation  \citep{2012MNRAS.422.1835S,2014ApJ...790...27L,2019MNRAS.485.3169P,2023MNRAS.526.1573S},
the authors concluded that quenched \hi-rich 
galaxies constitute only a tiny fraction of the general population of massive 
quiescent galaxies in the local Universe. This result is consistent with the 
widely-accepted fact that the cessation of star formation in a galaxy must be 
associated with the reduction of its cold gas reservoir \citep[see][for a recent 
review and references therein]{2022ARA&A..60..319S}.

Motivated by the observed 
relationship between the local surface density of star formation rate (SFR),
$\Sigma_{\rm SFR}$ and the local surface density of molecular gas 
$\Sigma_{\rm H2}$ down from kpc to sub-kpc scales \citep[e.g.][]{2008AJ....136.2846B,2011AJ....142...37S,2013AJ....146...19L, Lin2019, 2021MNRAS.502L...6E},
several physical models have assumed star formation to be regulated 
by the molecular fraction of the ISM, which is governed by its local conditions
such as hydrostatic mid-plane pressure or dynamical equilibrium pressure 
\citep[e.g.][]{1989ApJ...338..178E,1994ApJ...435L.121E,2002ApJ...569..157W,2004ApJ...612L..29B,2006ApJ...650..933B,2007ARA&A..45..565M,2010ApJ...721..975O,2022ApJ...936..137O, Ellison2024},
or the combination of local gas density and metallicity 
\citep[e.g.][]{Schaye2004,Krumholz2009,Krumholz2012}.
For rotationally supported disks, large-scale gravitational instabilities on 
scales larger than the ISM scale height have been assumed to play a 
major role in regulating star formation
\citep[e.g.][]{1960AnAp...23..979S,1964ApJ...139.1217T,1965MNRAS.130..125G,1984ApJ...276..127J,1992MNRAS.256..307R,1994ApJ...427..759W,1998ApJ...493..595H,2000ApJ...536..173T,2001MNRAS.323..445R,2002ApJ...577..206E,2007ARA&A..45..565M,2011ApJ...737...10E}.
In such models, the potential for 
gas clouds to collapse and form stars is set by processes that maintain a 
minimum effective velocity dispersion in the ISM, thus keeping the 
\citet{1964ApJ...139.1217T} parameter 
$Q\equiv \kappa c_{\rm s}/(\pi G \Sigma_{\rm gas})$ close to unity
(where $\kappa$ is epicyclic frequency, $c_{\rm s}$ is sound speed, 
and $\Sigma_{\rm gas}$ is the large-scale surface density of gas).
Whether star formation is driven by local conditions of the ISM or 
large-scale gravitational instabilities of the disk, its cessation process 
(quenching) must be somehow related to one of the two types of mechanisms, 
or both. In addition, environmental effects external to the host galaxy
should also be taken into account, as discussed above. 
 
Previous studies of galaxy quenching have been mostly limited to single-fiber 
spectroscopy or multiband photometry, thus probing either a limited region 
at galactic centers or the global properties of the whole galaxy. In order to 
have a complete picture of star formation cessation in galaxies, it is necessary 
to have resolved spectroscopy down to scales of molecular clouds which 
are believed to be the sites of star formation. 
Integral-field unit (IFU) surveys 
\citep{2001MNRAS.326...23B,2002MNRAS.329..513D,2010ApJ...716..198B,2011MNRAS.416.1680C,2012MNRAS.421..872C,2012A&A...538A...8S,2013AJ....145..138B,2014ApJ...796...52B,2014ApJ...795..158M,2015MNRAS.447.2857B,2015ApJ...798....7B} have provided 
resolved spectroscopy for star formation and quenching to be studied in 
large samples of galaxies down to kpc scales, though still much larger 
than the scales of individual star-forming regions and gas clouds 
(typically a few $\times10$ pc). Generally, quenched regions are identified 
as having no \ha\ emission, thus with substantially low surface brightness 
(\hasb) or equivalent width (\haew) in the \ha\ emission line. It is common  
to additionally require the quenched regions to not have 
experienced recent star formation over the past $\sim10^9$yr, thus 
dominated by old stellar populations. This requirement is formulated 
usually as \dnbreak$\ga 1.6$, where \dnbreak\ is the depth of the break 
at 4000\AA\ in the observed spectrum as defined by \cite{1999ApJ...527...54B}, 
known as a sensitive indicator of the average stellar age 
\citep{Kauffmann2003,2004MNRAS.351.1151B}.
Moreover, it has been increasingly suggested that quenched regions can 
be identified out of the low ionization (nuclear) emission line regions, or 
LI(N)ERs on the \citet[][BPT]{1981PASP...93....5B} diagnostic diagrams, which
can be regarded as regions where the star formation has already ceased, and
the ionization is primarily powered by hot evolved stars 
\citep[e.g.][]{Binette1994,Stasinska2008,Sarzi2010,2012ApJ...747...61Y,2013A&A...555L...1P,2013A&A...558A..43S,2016MNRAS.461.3111B,2016A&A...588A..68G,2017MNRAS.466.2570B,2017MNRAS.466.3217Z}.

In practice, previous studies have usually adopted two of the three selection 
criteria to define quenched regions, but with different pairwise combinations.
For instance, based on the Mapping Nearby Galaxies at Apache Point 
Observatory \citep[MaNGA][]{2015ApJ...798....7B} survey,
\citet{2015ApJ...804..125L} and \citet[][hereafter Paper I]{2018ApJ...856..137W}
identified quenched spaxels jointly by \haew$<2$\AA\ and \dnbreak$>1.6$, 
while \cite{Lin2019-quenching} selected LI(N)ERs with low \ha\ emission 
(\haew$<3$\AA). Also based on MaNGA data but differently from both of 
the above studies, \cite{Bluck2020-global-local,Bluck2020-central-satellite} firstly estimated SFR
surface densities (\Sigsfr) for both star-forming (SF) regions and non-SF 
regions (AGN, composite and low-S/N SF regions) as classified on the BPT 
diagram, using empirical relations of \Sigsfr\ with \hasb\ and \dnbreak\ 
respectively, and then selected quenched regions as those spaxels falling 
significantly below the ridge line of the resolved SF main sequence. 
As shown by the authors, the quenched regions so defined are essentially 
the non-SF regions on the BPT diagram with \dnbreak~$>1.45$. Scientifically,  
\citetalias{2018ApJ...856..137W} found that massive galaxies with 
stellar mass \mstar$\ga 10^{10}M_\odot$ quench their star formation 
from inside out. \citet{Lin2019-quenching} further found that the fraction 
of galaxies showing ``inside-out'' quenching increases with dark matter 
halo mass, and their results suggested that morphological quenching 
may be responsible for the inside-out quenching in galaxies of all 
environments. \citet{Bluck2020-global-local,Bluck2020-central-satellite} trained a multilayered 
artificial neural network (ANN) and a random forest (RF) to classify 
spaxels into SF and quenched regions given various local/global 
parameters, finding central velocity dispersion is the best single parameter
in predicting quenching in central galaxies. This finding was interpreted 
by the authors as the observational evidence for AGN feedback. 

In this work, we will adopt the 
same quenching definition as in \citetalias{2018ApJ...856..137W} and 
examine the significance of global and resolved properties to quenching 
in a similar way to \citet{Bluck2020-global-local,Bluck2020-central-satellite}.
In particular, we will make comparisons for subsamples of gas-rich 
quenched regions (GRQRs) and gas-poor quenched regions (GPQRs) 
selected by the surface density of cold gas (\Siggas). We will also perform 
this analysis but adopting the same 
quenching definitions as in both \citet{Lin2019-quenching} and 
\citet{Bluck2020-global-local,Bluck2020-central-satellite}, in order to examine the possible 
dependence of our results on quenching definition. As we will show, 
quenched regions selected by different criteria indeed present different 
correlations with global and local properties, which lead to different 
implications and interpretations on the underlying quenching mechanisms. 
The dependence on quenching definition has been largely overlooked 
in previous studies, and needs to be taken into account in future 
studies. 

This paper is organized as follows. In \autoref{sec:DandM} we describe 
the data and parameter measurements used in this work. We then present 
our results in \autoref{sec:Res}. We discuss our results in \autoref{sec:discussion}
and summarize in \autoref{sec:Conc}. Throughout this paper we assume a 
flat $\Lambda$CDM cosmology with parameters given by the 
WMAP nine-year results: $\Omega_{m}=0.286, \Omega_{\Lambda}=0.714$ 
and $H_{0}=69.3 \mathrm{km} \mathrm{s}^{-1} \mathrm{Mpc}^{-1}$ 
\citep{2013ApJS..208...19H}.

\section{Data and Measurements} \label{sec:DandM}

\subsection{MaNGA} \label{sec:manga}

MaNGA is one of the three core programs of the fourth-generation Sloan Digital 
Sky Survey \citep[SDSS-IV;][]{2017AJ....154...28B}, and is the largest integral field
spectroscopy (IFS) survey ever accomplished \citep{2015ApJ...798....7B}.
During its 6-year operation form July 2014 through August 2020, MaNGA obtained IFS
data for $\sim$10,000 galaxies in the nearby Universe. MaNGA utilized 29 integral
field units (IFUs), including 17 hexagonal science IFUs with a field of view
(FoV) ranging from $12''$ to $32''$ and 12 seven-fiber mini-IFUs for flux
calibration \citep{2015AJ....149...77D}. The IFU fibers are fed to the two
dual-channel BOSS spectrographs \citep{2013AJ....146...32S} on the
Sloan 2.5-m telescope at the Apache Point Observatory \citep{2006AJ....131.2332G}
to produce MaNGA spectra with a spectral resolution of $R \sim 2000$ over
the wavelength range from 3600\AA\ to 10300\AA. The spectra reach
an $r$-band signal-to-noise ratio (SNR) of 4-8 per \AA\ per $2''$ fiber at
1-2 effective radius ($R_e$) of the target galaxies, with a typical exposure
time of three hours.

The targets of MaNGA are selected from the NASA Sloan Atlas
\citep[NSA;][]{Blanton2011}, and are divided into three samples:
Primary, Secondary and Color-Enhanced samples. By selection the Primary
and Secondary samples have a flat distribution of $K$-corrected $i$-band
absolute magnitude $M_i$, and the targets of the two samples are observed
with IFUs covering out to 1.5$R_e$ and 2.5$R_e$ respectively.
The Color Enhanced sample further selects galaxies on the $NUV-r$ versus
$M_i$ diagram that are not well sampled by the Primary/Secondary samples.
Overall, MaNGA samples span a wide stellar mass range of
$5 \times 10^8 M_\odot h^{-2} \le M_\ast \le 3 \times 10^{11} M_\odot h^{-2}$
and a redshift range of $0.01 < z < 0.15$ with a median redshift of $z\sim0.03$
\citep{2017AJ....154...86W}.

The raw data of MaNGA are reduced with the Data Reduction Pipeline
(DRP; \citealt{2016AJ....152...83L}) to produce a datacube for each target,
with a spaxel size of $0^{\prime\prime}.5\times0^{\prime\prime}.5$ and
an effective spatial resolution that can be described by a Gaussian with
a full width at half maximum (FWHM) of $\sim 2^{\prime\prime}.5$.
The absolute flux calibration of the spectra is better than 5\% for more
than 80\% of the wavelength range. More details about the flux calibration,
survey execution strategy and data quality tests can be found in
\citet{2016AJ....152..197Y} and \citet{2016AJ....151....8Y}. In addition,
by performing full spectral fitting to the datacubes from DRP, the Data
Analysis Pipeline \citep[DAP;][]{2019AJ....158..231W,2019AJ....158..160B}
derives measurements of stellar kinematics, emission lines and spectral
indices from each spectrum.
All the MaNGA data including the DRP and DAP products of the full
sample of 10,010 galaxies are released as a part of the final data release
of SDSS-IV \citep[DR17;][]{SDSS_DR17}.

\subsection{Ancillary data and spaxel measurements}
\label{sec:measurement}

In addition to MaNGA data, we use ancillary data from a number of 
previous studies: (1) galaxy morphology type T-type
from \cite{2018MNRAS.476.3661D}, (2) galaxy photometric parameters
such as bulge-to-total ratio $B/T$ and bulge effective radius
$R_{e,\mathrm{bulge}}$ from \cite{2016MNRAS.455.2440M},
(3) position angle (PA) and axis ratio of galaxy image
from NSA, and (4) galaxy interaction/merger identification from
\cite{2018ApJ...868..132P}. The last one is available only for an earlier
sample of MaNGA, the MaNGA Product Launch 6 (MPL6), which
includes 4690 galaxies, a random subset of the final MaNGA sample.
Therefore, we restrict ourselves to MPL6 and select our sample out of it.
As the first step, we exclude 1343 galaxies that are not included, 
or are flagged in any of the abovementioned catalogs. Next, we 
remove 806 interacting/merging galaxies, which are flagged 
{\tt\string flag\_pair > 0} or {\tt\string p\_merger > 0.4}
by \cite{2018ApJ...868..132P}, as well as 850 elliptical galaxies 
with T-type $<0$ or $B/T=1$. Finally, four galaxies flagged in 
the MaNGA datacube are also discarded. These restrictions result in 
a sample of 1687 galaxies. We call this sample as ``starting sample''.

We make use of a variety of parameters for each spaxel 
in our galaxies, which were measured and used in our previous studies,
and are made publicly available in \cite{2023ChPhB..32c9801L}. Here 
we briefly describe how the parameters were measured and refer the 
reader to \cite{2023ChPhB..32c9801L} for more details. 
First of all, for kinematics, we use the MaNGA DAP products, including
stellar velocity (\vstar) and velocity dispersion (\sigstar), gas
velocity and velocity dispersion as measured from different emission lines,
and flux of the emission lines. For each spaxel, we then estimate
the gas-phase color excess \ebv\ from Balmer decrement, with which
we correct the effect of dust extinction on the emission line fluxes.
We then estimate the surface density of star formation rate (SFR)
\Sigsfr\ from \Sigha\
(the extinction-corrected H$\alpha$ surface brightness), according
to the estimator from \citet{1998ARA&A..36..189K} with projection
effect corrected. The method is principally valid for regions predominantly ionized by newly formed OB stars, which represent the majority of our sample. However, in regions where other ionization sources cannot be ignored, especially the quenched regions we will select in the next section, the \Sigsfr\ derived using this method can only be considered as an upper limit.

Next, for a given spaxel, we apply the technique of \cite{2020ApJ...896...38L}
to the observed MaNGA spectrum to estimate a model-independent
attenuation curve, and we correct the effect of stellar motions and dust
attenuation for the spectrum using \vstar, \sigstar\ and the
attenuation curve. We then measure stellar population parameters,
e.g. stellar surface mass density \Sigstar, as well as
spectral indices, e.g. \ha\ equivalent width \haew\
and the 4000\AA\ break \dnbreak, by performing full spectral
fitting to the dust-free spectrum with zero velocity broadening.
Details of the technique of measuring dust attenuation curves and full
spectral fitting can be found in \cite{2020ApJ...896...38L, Li2021}.

In addition, we have estimated the surface density of both molecular gas 
\Sigmol\ and Hydrogen atomic gas \SigHI\ for each spaxel. We briefly describe 
the two parameters and refer the reader to Appendix \ref{sec:H2_measure} for details.
For \Sigmol, due to the lack of spatially resolved CO observations with 
resolution comparable to MaNGA, we have developed an empirical estimator
which uses a combination of four parameters from MaNGA to predict the 
fraction of molecular gas mass density 
$f_{\mathrm{H}_2}\equiv$~\Sigmol/(\Sigstar+\Sigmol) for each spaxel.
We calibrate and test the estimator using the EDGE-CALIFA survey which
provides both IFS and CO mapping for a sample of nearby galaxies.
We show that the estimator provides unbiased estimation of \Sigmol\
with a reasonably small scatter of 0.26 dex (Appendix \ref{sec:H2_measure}).
Similarly, \SigHI\ is also estimated empirically for the lack of spatially resolved 
H{\sc i} observations. We have considered three different methods,  
finding similar \SigHI\ measurements. More importantly, no matter what 
method is adopted, \SigHI\ contributes only a small fraction to the total gas 
mass density which is dominated by \Sigmol\ in most cases. Therefore, 
for simplicity we assume a constant H{\sc i} surface mass density for all the 
spaxels, \SigHI~$= 6 \times 10^6 {\mathrm{M}}_\odot {\mathrm{kpc}}^{-2}$,
following \citet{2020MNRAS.492.2651B} (Appendix \ref{sec:H2_measure}).

\begin{figure*}
	%	\plotone{GRQRs.pdf}
	\centering
	\includegraphics[width=\textwidth]{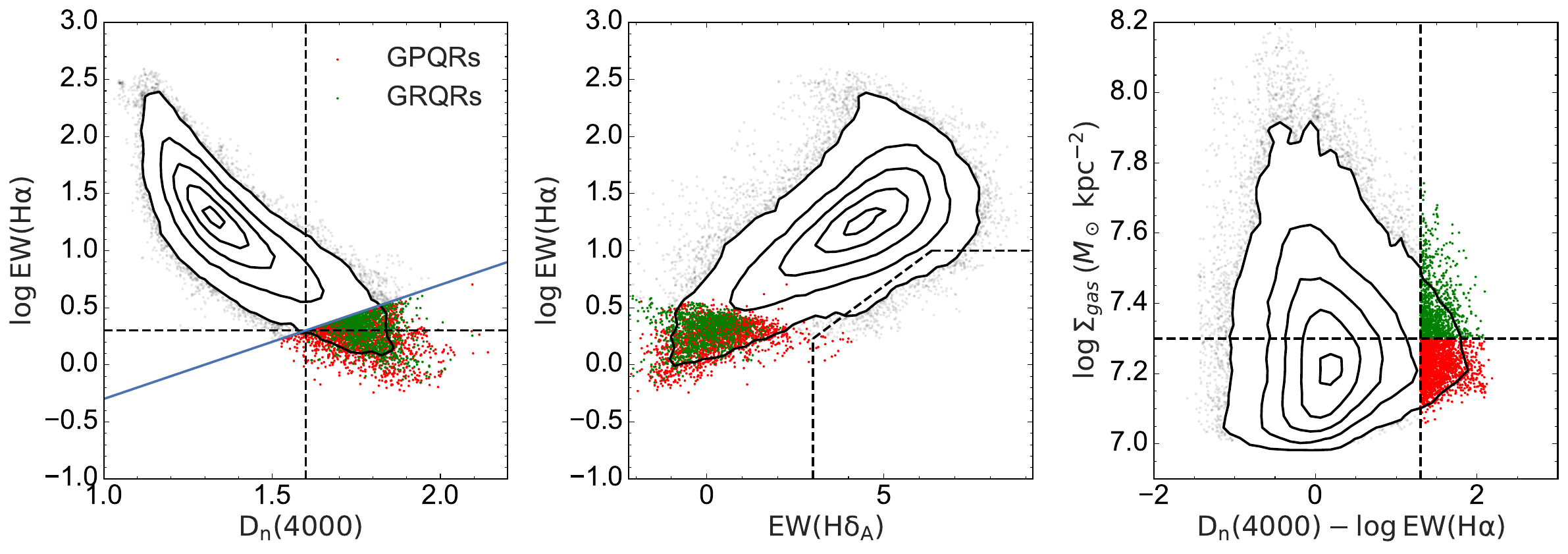}
	\caption{Left panel: The $\log$ \haew\ versus \dnbreak\ diagram of our sample. Contour lines represent data percentages (10\%, 40\%, 70\%, 90\%, and 99\%), while black dots indicate data outside the 99\% contour line. GRQRs and GPQRs are denoted by green and red dots, respectively (same for other two panels). The horizontal and vertical dash line represent \haew\ $= 2 \AA$ and \dnbreak\ $= 1.6$ separately.
		Center panel: The \haew\ versus absorption \hdew\ diagram of our sample. The right-bottom corner are post-starburst region according to \cite{2019MNRAS.489.5709C}.
		Right panel: The $\log\ \Sigma_{gas}$ versus \Qtracer\ diagram of our sample. The horizontal dash line represents the boundary between gas-rich and gas-poor regions at $\log\ \Sigma_{gas} (M_\odot \mathrm{kpc}^{-2}) = 7.3$. Spaxels on the right side of the vertical dash line (\Qtracer\ $> 1.6 - \log2 \approx 1.3$) are considered as quenched spaxels, which would be found in the lower right side of blue solid line in left panel. \label{fig:GRQRs}}
\end{figure*}

\subsection{Selection of spaxels}\label{sec:sample_selection}

Based on the spaxel measurements, we select a sample of spaxels for 
the following analysis. First, we require the selected spaxels to 
have SNR$>3$ in all the following parameters: 
\sigstar, $v_{H\alpha}$ (\ha\ gas velocity),
$\sigma_{H\alpha}$ (\ha\ gas velocity dispersion),
\ha\ flux, \hb\ flux, \nii$\lambda$6585 flux,  \dnbreak\ and \haew.
By implementing a stringent SNR cut, we aim to minimize any potential bias and ensure that the comparison between local and global properties is done on a more equal and reliable basis. The global properties of MaNGA galaxies are typically measured with high SNR. If local parameters have lower SNR, it could lead to an underestimation of the correlation between these low SNR parameters and the properties of interest when compared to the global properties.
In addition, to ensure reliable measurements
of the gas mass surface density (see Appendix \ref{sec:H2_measure} for details),
each spaxel has to also meet the following requirements in order to be included:  
gas-phase color excess $0 < $~\ebv~$< 1$,
molecular gas mass surface density $6.6<\log$(\Sigmol/$\mathrm{M}_\odot\mathrm{kpc}^{-2}$)~$<8$,
the \nii-to-H$\alpha$ line ratio $-3<\log($\nii/\ha$)<1$,
the galactic-centric radius within twice the effect radius 
($R<2R_e$) and beyond three times the bulge radius 
($R\geq 3R_{e,\mathrm{bulge}}$). 
With the last criterion we consider only the spaxels in galactic discs.
Spaxels within $30^{\circ}$ of the minor axis are excluded due to 
potential measurement inaccuracies in rotation velocity and increased 
risk of contamination by extraplanar emission.
Galaxies that lack spaxels meeting these requirements are 
automatically excluded from our sample.
%We have further excluded 150 galaxies each with lessn than five 
%spaxels with measurements of $\kappa$ and $Q$, 
We have further excluded 21 galaxies whose \ha\ orientation differs significantly 
from their stellar orientation. For this purpose, we have compared 
the isophotes of $\Sigma_{H\alpha}$ with the ellipse of the $r$-band 
light distribution as determined from the position angle and axis ratio 
from NSA. The star-forming regions in such ``misaligned'' galaxies 
may be dynamically unstable. 

Our final sample includes a total of 265,304 spaxels, 
distributed in 1205 galaxies. This sample will be referred to as 
our ``parent sample'' in what follows.

% Here, when calculating SNRs, 
% the noise is given by the geometric mean of two errors: one is the 
% observational error from the observed spectrum and one is the uncertainty 
% of spectral fitting. 

\begin{figure*}
	%	\plotone{KSlaw.pdf}
	\centering
	\includegraphics[width=\textwidth]{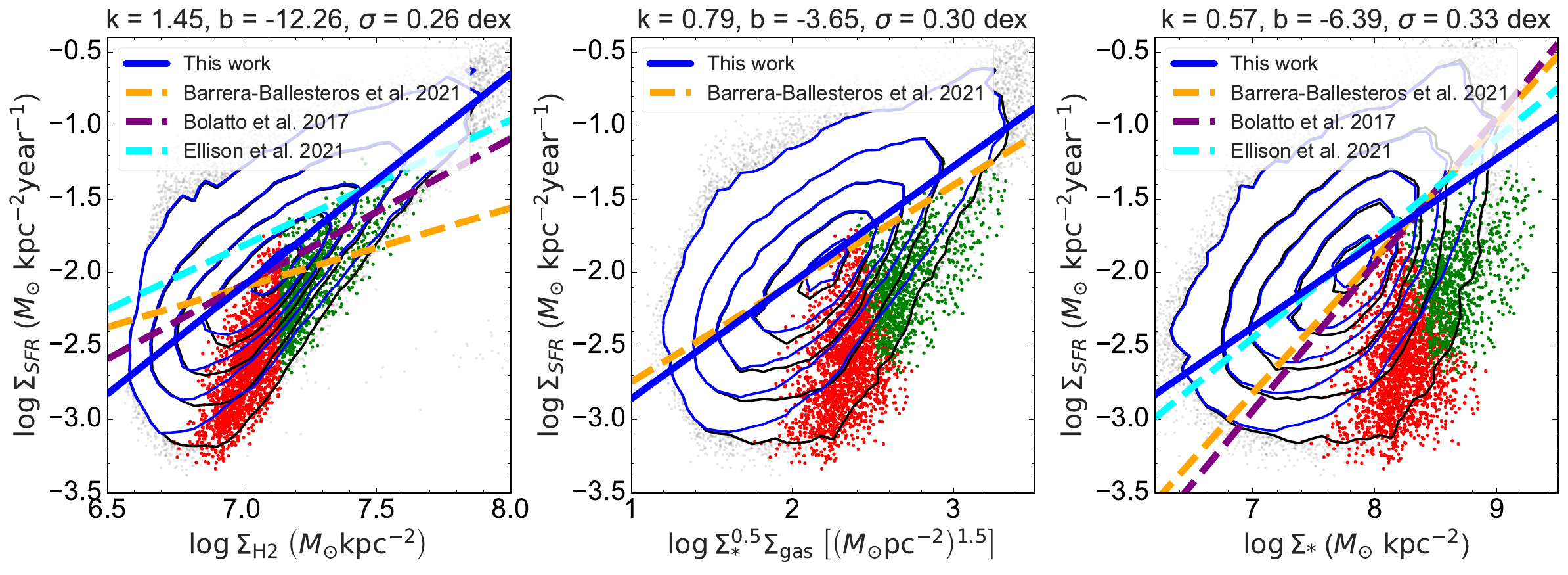}
	\caption{The star formation scaling laws in our sample. From left to right, the three panels show rSK law, extend rSK law and rSFMS, separately. The meanings of green dots and red dots are same as \autoref{fig:GRQRs}. The blue and black contours display the distribution of star-forming regions and all regions in our sample, with each line indicating specific data percentages (10\%, 40\%, 70\%, 90\%, and 99\%). Additionally, blue and black dots mark data points outside the corresponding 99\% contour line. The blue solid lines represent the best-fitted scaling law on the star-forming regions in our sample, while colorful dash lines are scaling laws reported in literatures. Each relation have been converted to the \cite{2003PASP..115..763C} initial mass function case. \label{fig:KSlaw}}
\end{figure*}

\begin{figure*}
	\centering
	\includegraphics[width=\textwidth]{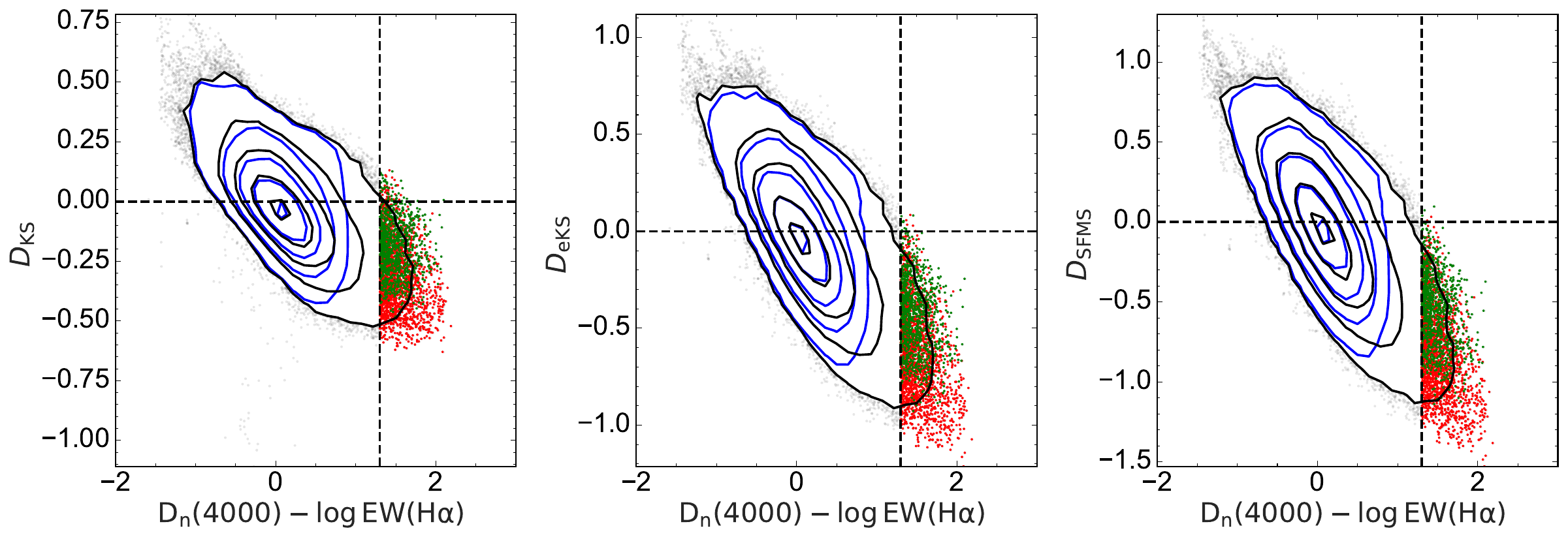}
	\caption{The distance to scaling law versus \Qtracer\ diagram for all spaxels in our sample. From left to right, the three panels show the distance to rSK law, extend rSK law and rSFMS, separately. The meanings of contours and dots are same as \autoref{fig:KSlaw}. \label{fig:DKSlaw}}
\end{figure*}

\section{Results} \label{sec:Res}

\subsection{Identification of quenched but gas-rich regions} \label{sec:GRQRs}

We firstly identify quenched regions in our galaxies by selecting
spaxels with no/weak ongoing star formation and dominated by
old stellar populations. In practice, we jointly use \dnbreak\ and
\haew\ following \citetalias{2018ApJ...856..137W}, where a spaxel
was classified to be a quenched region if \dnbreak~$\ge 1.6$ 
(dominated by old populations) and \haew~$\le 2$\AA\ (consistent with the typical \haew\ ionized by old population; e.g. \citealt{Binette1994,Stasinska2008,Sarzi2010,2012ApJ...747...61Y,2013A&A...555L...1P,2013A&A...558A..43S,Sanchez2014,2016MNRAS.461.3111B,2016A&A...588A..68G,2017MNRAS.466.2570B,Morisset2016,2017MNRAS.472.4382R,2017MNRAS.466.3217Z}). Here, for simplicity, we combine the two 
criteria into a single one:
\Qtracer\ $\ge 1.6-\log{2}=1.3$. 
The left panel in \autoref{fig:GRQRs} displays the distribution of 
all the spaxels from our parent sample in the diagram of 
$\log$\haew\ versus \dnbreak,
with the new divider indicated by the blue line. The spaxels identified as quenched fall below this line.
The horizontal and vertical dashed lines
indicate the two single-parameter cuts. As can be seen,
the new criterion selects some extra spaxels with \haew\ slightly
larger than 2\AA\ or \dnbreak\ slightly smaller than 1.6. In the center
panel of the same figure we show the distribution of our sample and
the quenched spaxels in the diagram of \haew\ versus \hdew,
equivalent width of the \hd\ absorption line. This diagram is commonly
used to select post-starburst galaxies/regions, which are expected
to fall in the area enclosed by the black dashed lines. The quenched
spaxels are located in the lower-left tail of the whole sample,
with low values in both parameters indicative of weak star formation
and old populations. With only a few exceptions, all the spaxels
are far away from the post-starburst regime. This implies that
these regions have long been quenched, with no strong star formation
in the past 1-2 Gyr. We notice that the extra spaxels with \haew\
slightly higher than 2\AA\ also have small values in \hdew\ comparable
to the spaxels of \haew$<2$\AA.

The rightmost panel of \autoref{fig:GRQRs} displays the diagram of
\Siggas$\equiv$\SigHI+\Sigmol\
versus \dnbreak-$\log$\haew, where we divide the quenched
spaxels into two subsamples with  $\log($\Siggas$/\mathrm{M}_\odot\mathrm{kpc}^{-2})\ge 7.3$
and $<7.3$ respectively. This dividing cut is slightly larger than
the peak of the full sample which is at
$\log($\Siggas$/\mathrm{M}_\odot\mathrm{kpc}^{-2})\sim7.2$.
The two subsamples consist
of 1060 gas-rich quenched regions (GRQRs hereafter) and
2069 gas-poor quenched regions (GPQRs hereafter), respectively.
As can be seen from the left two panels in \autoref{fig:GRQRs},
the two types of quenched spaxels present similar distributions,
indicating similar star formation status.
The main purpose of this work is to understand the reason why
the two types of regions are both quenched but with so different gas content.  
We note that the results of our work are robust to the dividing cut
which is arbitrarily set at
$\log($\Siggas$/\mathrm{M}_\odot\mathrm{kpc}^{-2})=7.3$.
We have repeated our analysis by using more extreme selection criteria,
with $\log($\Siggas$/\mathrm{M}_\odot\mathrm{kpc}^{-2})> 7.4$
for GRQRs and
$\log($\Siggas$/\mathrm{M}_\odot\mathrm{kpc}^{-2})< 7.2$
for GPQRs, finding our main results to remain similar but become
more noisy due to smaller sample sizes.

\begin{figure*}
	\centering
	\includegraphics[width=\textwidth]{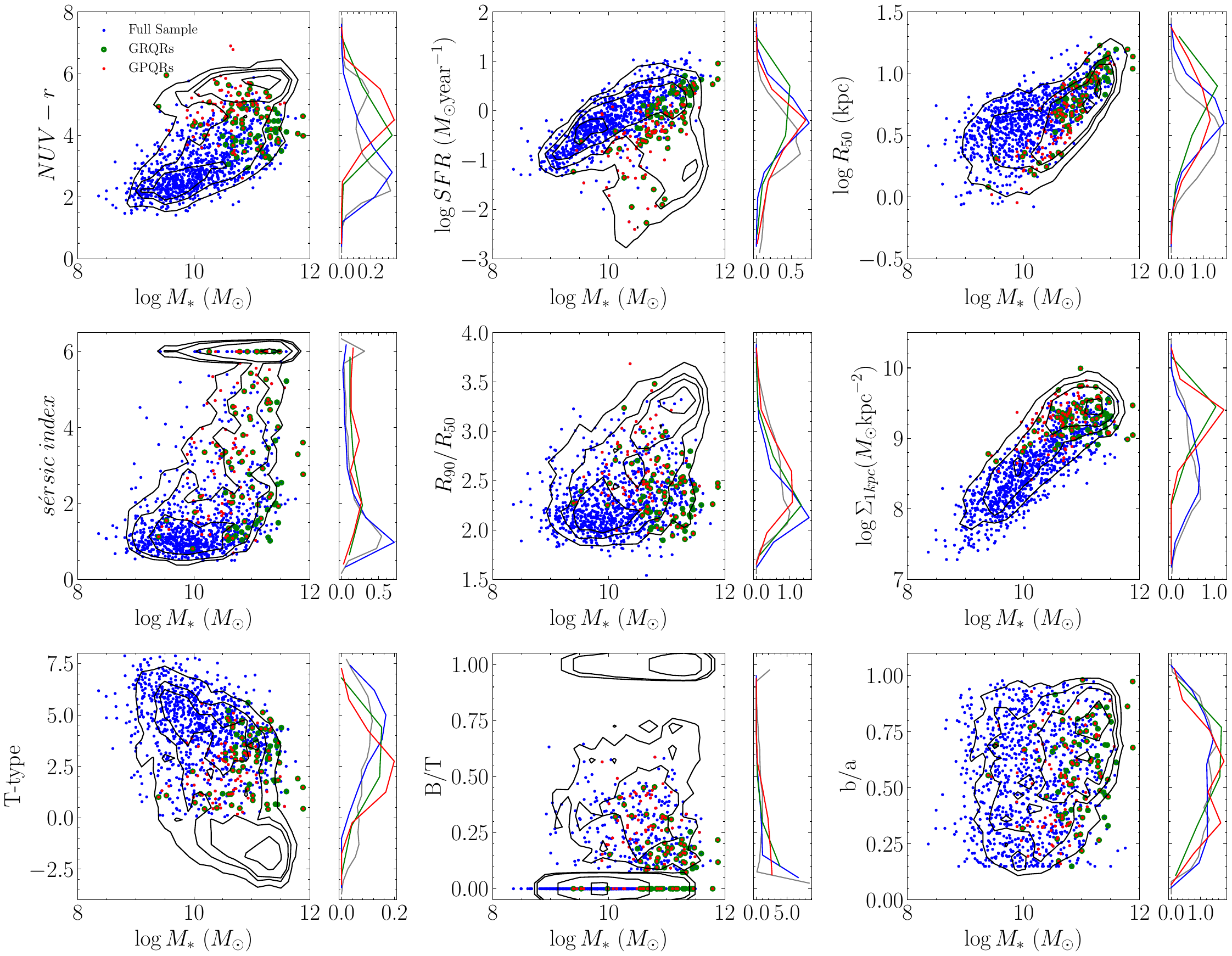}
	\caption{ Comparison of global properties among all MaNGA galaxies (background contour, the contour lines include 10\%, 40\%, 70\% and 90\% data separately), galaxies in our sample (blue dots), the host galaxies of GRQRs (green ring) and the host galaxies of GPQRs (red dots). The distribution of MaNGA galaxies are corrected by galaxy weight. Each panel is a y-axis parameter versus galaxy mass diagram. The y-axis parameters of these nine panels are color (NUV - r), logarithmic star formation rate, galaxy size, \sersic\ index, Concentration ($r_{90}/r_{50}$), stellar mass surface density in central 1 kpc, T-type, B/T, and axial ratio (b/a), from left to right, and top to bottom. The sub-panel in right side of each panel is the probability density profile of y-axis parameter. Note that for the host galaxies of GPQRs, the red dots cover the blue dots, which leads to these dots show as a red one.  \label{fig:compare_global}}
	% The classification of the host galaxies of GRQRs and the host galaxies of GPQRs is not mutually exclusive. And both the host galaxies of GRQRs and GPQRs are included in the full sample.
	% , where the MaNGA galaxies, galaxies in our sample, the host galaxies of GRQRs and the host galaxies of GPQRs is colored by gray, blue, green and red separately
	% For the incompleteness of morphology measurement (\sersic\ index, concentration, T-type and B/T) of MaNGAa MPL-8 galaxies, the data on these panels are galaxies with corresponding parameters.
\end{figure*}

In \autoref{fig:KSlaw} we further examine the distribution of the
GRQRs and GPQRs on three star-forming scaling laws (panels from left to right):
\Sigsfr\ versus \Sigmol\ (resolved Schmidt-Kennicutt law, or rSK law;
\citealt{1959ApJ...129..243S, 1963ApJ...137..758S, 1998ApJ...498..541K}),
\Sigsfr\ versus $\rm \Sigma_\ast^{0.5}\Sigma_{gas}$
(extended rSK law, as proposed by
\citealt{2011ApJ...733...87S, 2018ApJ...853..149S}),
and \Sigsfr\ versus \Sigstar\ (resolved star-forming main sequence, rSFMS).
In each panel, the red and green dots display the GPQRs and GRQRs, while the black contours show the distribution of all the spaxels from the parent sample. These scaling laws have been obtained in previous studies at spatial resolutions comparable to or better than those of MaNGA \citep{2017ApJ...846..159B, Lin2019, Sanchez2020, 2021arXiv210102711B, Ellison2021_nonuni, Sanchez2021, Sanchez2021_global_to_resolved, Pessa2021, Baker2022, Lin2022, Sanchez2023}. For comparison, plotted as colored dashed lines in \autoref{fig:KSlaw} are
	the average scaling relations of spatially resolved regions in nearby galaxies, as obtained by previous studies based on MaNGA \citep{2021arXiv210102711B}, EDGE-CALIFA \citep{2017ApJ...846..159B} and ALMaQUEST \citep{Ellison2021_nonuni}.
Since these relations are usually identified using star-forming relations, for a fair comparison we have specifically selected the star-forming regions from our parent sample by requiring them to have \haew\ $>$ 6 \citep{CidFernandes2010, Sanchez2014, Morisset2016, Sanchez2021_global_to_resolved} and fall below the \cite{Kauffmann2003} line in the BPT diagram. The distribution of this sample is shown as blue contours in the same figure, and the best-fitted linear relation of the star-forming regions, as obtained by applying the Ordinary Least Squares (OLS) algorithm, is plotted as the blue solid line in each panel. The best-fit parameters (slope $k$, intercept $b$, scatter $\sigma$) are indicated above each panel. Overall, both the best-fit relation and the scatter from our sample are consistent with one or multiple of the previous studies. Differences between different studies may be caused by different sample selections, different ways of estimating gas mass fractions, and different fitting algorithms.
%	 It is observed that the rSK law derived in our sample exhibits a deviation from the correlation reported in the literature. Nevertheless, this deviation is comparable to the differences within the calibration reported by different works. This difference could potentially be attributed to the different methods used to derive $\Sigma_{H2}$: we use estimator described in \autoref{sec:H2_measure}, while \cite{2021arXiv210102711B} utilize an $A_V$-based estimator, and \cite{2017ApJ...846..159B} and \cite{Ellison2021_nonuni} rely on CO-derived $\Sigma_{H2}$. Our results are in agreement with the findings of \cite{2021arXiv210102711B} for the extended rSK law. For rSFMS, we find a very similar slope and zero point as reported by \cite{Ellison2021_nonuni}. Employing Orthogonal Distance Regression (ODR) would result in a larger slope, very similar to other two works. This suggests that the slope of the relation can be influenced by the fitting method used, as shown in \cite{Ellison2021_nonuni}. With regard to the scatter of the scaling relation, we find that the scatter of our scaling law (0.25 dex for rSK, 0.33 dex for rSFMS) is comparable to that reported by \cite{Ellison2021_nonuni} (0.28 dex for rSK, 0.39 dex for rSFMS) based on the same fitting method. 

As can be seen from \autoref{fig:KSlaw}, the majority of the quenched spaxels including both GRQRs and GPQRs fall below the average relations in all the panels. The fall of quenched regions below the rSFMS suggests a reduction in specific star formation rate (sSFR), while the fall below the rSK law suggests a reduction in star formation efficiency (SFE).
This behaviour is more clearly seen from \autoref{fig:DKSlaw}, where we 
show the deviation of spaxels from the average relations
as functions of \dnbreak-$\log$\haew, the single parameter 
adopted above for the selection of quenched regions. 
This parameter shows an anti-correlation with the deviation in all 
the panels. In particular, we find most of the quenched regions defined 
by this parameter have negative deviations, while those with 
negative deviations span a wide range in this parameter, i.e. 
not all of them are quenched. In other words, to fall below 
the star-forming scaling relations is a necessary condition, but not a 
sufficient condition for a spaxel to be considered as a quenched region. 
This is because our selection criteria require the quenched regions 
to be not only weak in ongoing star formation, but also dominated by
old stellar populations.

%94\% of GRQRs in galaxies with HI observation are also selected 
%as GRQRs when the $\Sigma_{HI}$ is estimated by \cite{2016MNRAS.460.2143W, 2020ApJ...890...63W}'s model. As for molecular gas, when fixed HI as the constant, 70\% of predicted gas rich regions ($\log{\Sigma_{gas}} > 7.3$) in EDGE-CALIFA are actual gas rich. 

\begin{figure*}
	\centering
	\includegraphics[width=\textwidth]{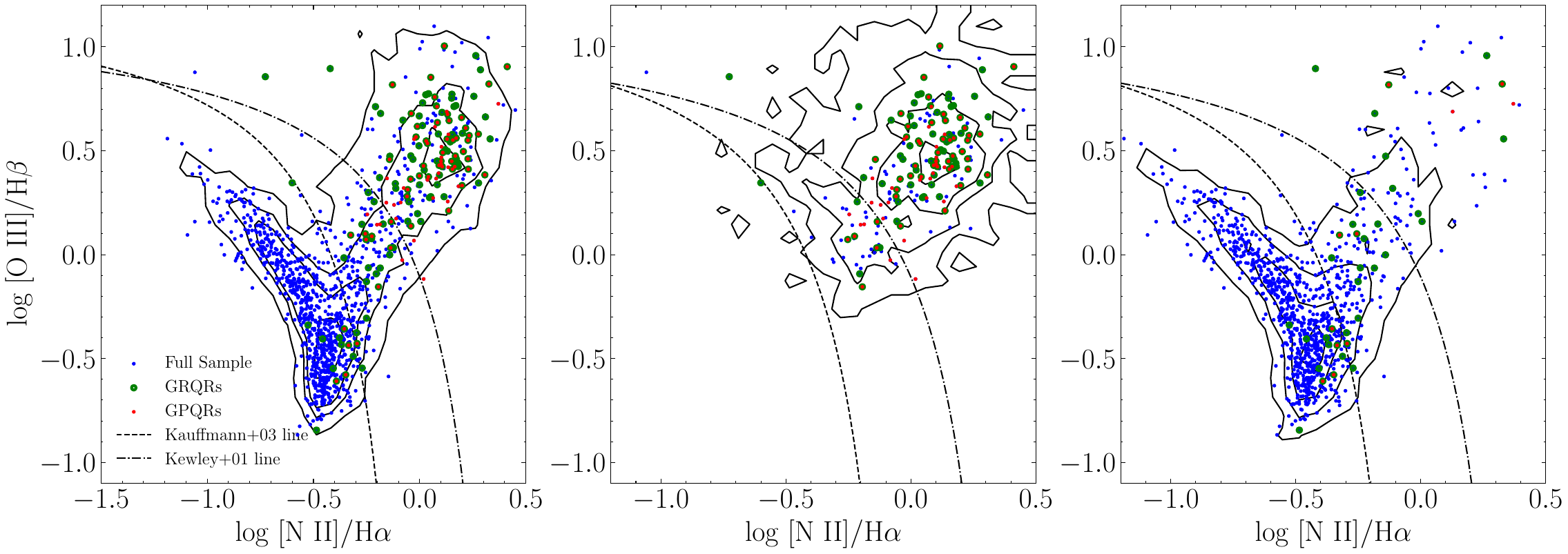}
	\caption{BPT diagnostics of our sample. We use the line ratios of central spaxels to trace the existence of AGN for MaNGA galaxies (background contour, the contour lines include 10\%, 40\%, 70\% and 90\% data separately), galaxies in our sample (blue dots), the host galaxies of GRQRs (green ring) and the host galaxies of GPQRs (red dots). The distribution of MaNGA galaxies have been corrected by the galaxy weight. The center and right panel is same as left panels but for retired galaxies (central \haew  $\ <$ 3 \AA) and star forming galaxies (central \haew $\ >$ 3 \AA), separately. \label{fig:AGN_diagnose}}
\end{figure*}

\subsection{Global properties of host galaxies} \label{sec:glo_loc_prop}

In \autoref{fig:compare_global} we examine the global properties of
the host galaxies of both GRQRs and GPQRs. The properties considered
include color index $NUV-r$, the total SFR, the $r$-band effective
radius $R_{50}$, \sersic\ index $n$, concentration index $R_{90}/R_{50}$,
the surface stellar mass density within the central 1kpc $\Sigma_{\rm 1kpc}$,
morphology as quantified by T-type, the bulge-to-total luminosity ratio
$B/T$, and the minor-to-major axis ratio $b/a$ as measured
from $r$-band images. Here, $R_{50}$ and $R_{90}$ are
the radii enclosing 50\% and 90\% of the total light in $r$-band.
Among these properties, total SFRs and stellar mass are taken from 
GSWLC \citep{GLWSC1, GLWSC2}, T-types are from
\citet{2018MNRAS.476.3661D},
$B/T$ from \citet{2016MNRAS.455.2440M}, $\Sigma_{\rm 1kpc}$
are measured by ourselves from the MaNGA data, and all the rest
properties are provided in NSA.

In the figure the host galaxies of GRQRs and GPQRs are plotted as 
green and red dots.
For comparison, we plot the distribution of all the galaxies in the parent 
sample (blue dots) from which the quenched regions are selected,
as well as the distribution of the full sample of MaNGA galaxies
from SDSS/DR17 as the background contours, for which we have
corrected the incompleteness due to MaNGA sample
selection using the weights provided by \cite{2017AJ....154...86W}.
By selection, our parent sample includes only late-type galaxies
(T-type$\ge0$ and $B/T<1$) with no signatures of interaction/merger.
As expected, this sample is dominated by globally blue and
star-forming galaxies with relatively large size and low concentration,
\sersic\ index and $B/T$ at fixed mass. The quenched
regions including both GRQRs and GPQRs tend to be hosted by
massive galaxies (\mstar$\ga 10^{10}\rm M_\odot$) with relatively
red colors ($NUV-r\ga 3$), low SFR and high central density at fixed mass,
but spanning wide ranges in other parameters that are similar
to the parent sample.
We note that our sample spans a full range of $b/a$, indicating
that our sample selection is not biased by dust extinction effect.
It is interesting to note that the GRQRs and GPQRs are similar
in all the host galaxy properties considered, and that most of the
host galaxies of GRQRs also host GPQRs at the same time. This
strongly implies that the conditions responsible for GRQRs/GPQRs
must be largely independent on the global properties of their host galaxies.
Rather, local processes on scales of the MaNGA spaxels or even
smaller scales are more likely to be at work.

\autoref{fig:AGN_diagnose} displays the BPT diagram \citep{1981PASP...93....5B}
for the host galaxies of GRQRs and GPQRs, compared to the parent sample
and the full MaNGA sample.
We use the emission line measurements from the central
spaxels in MaNGA datacubes.
We use the empirical criteria from \citet{2001ApJ...556..121K}
and \citet{Kauffmann2003} to divide the diagram into three
regimes dominated by different ionizing sources.
The left panel show the results for all the host galaxies as a whole, while the center and right panels show the results separately for the two subsets of host galaxies with central \haew\ values being either $<$ 3\AA\ or $\ge$ 3\AA. This division is suggested in the literature (e.g. \citealt{CidFernandes2010, CidFernandes2011, Sanchez2014, Sanchez2018, Sanchez2020, Sanchez2021_global_to_resolved}) to mitigate the confusion arising from different ionization conditions based solely on the location in the BPT diagram. Although inaccurate for individual sources, this \haew\ threshold statistically distinguishes the ionization dominated by AGN (\haew\ $\ge$ 3\AA) from that dominated by low mass evolved stars (\haew\ $<$ 3\AA). 
Overall, the galaxies hosting GRQRs and GPQRs exhibit similar distributions in the BPT diagram, predominantly located above the line defined by \cite{2001ApJ...556..121K}. These galaxies mainly have \haew\ $<$ 3\AA. 
The fraction of the quenched region host galaxies that satisfy \haew\ $\ge$ 3\AA\ and locate above the line defined by \cite{Kauffmann2003} is limited.
In addition to the \haew, the central H$\alpha$ velocity dispersion $\sigma_{\rm H\alpha, c}$ can also provide complementary information for ionization condition classification. By following the criteria \haew\ $>$ 3\AA\ and $\sigma_{\rm H\alpha, c}$ $>$ 57 km s$^{-1}$ as suggested by \cite{Sanchez2024} for AGN selection, the AGN fraction in the host galaxies of quenched regions is approximately 20\%. These results suggest that the majority of host galaxies of both types of quenched regions are not obviously AGN host galaxies. The similarity between the GRQRs and GPQRs again indicates weak or no dependence of quenching on the global properties of host galaxies.

\subsection{Significance of resolved properties to quenching}\label{sec:resolved_properties}

\begin{figure}
	\includegraphics[width=0.48\textwidth]{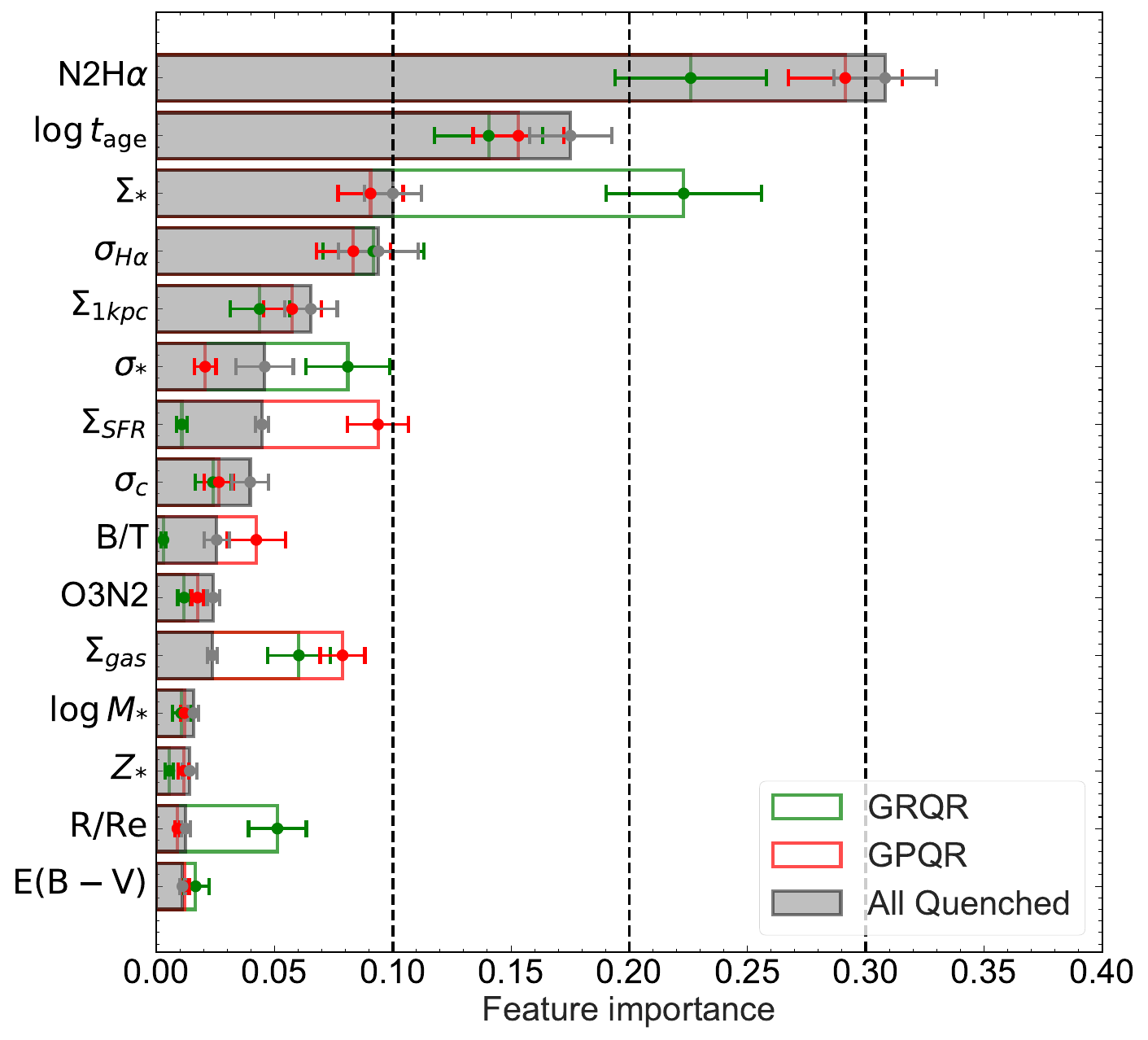}
	\caption{The Gini feature importance for selecting quenched regions (gray), GPQRs (red), and GRQRs (green) of each input property. The error bar shows 1 $\sigma$ uncertainty of importance, estimated by re-selecting the subsample of non-quenched spaxels and re-splitting the train/test set. \label{fig:RF_importance}}
\end{figure}

\begin{figure*}[ht!]
	\centering
	\includegraphics[width=\textwidth]{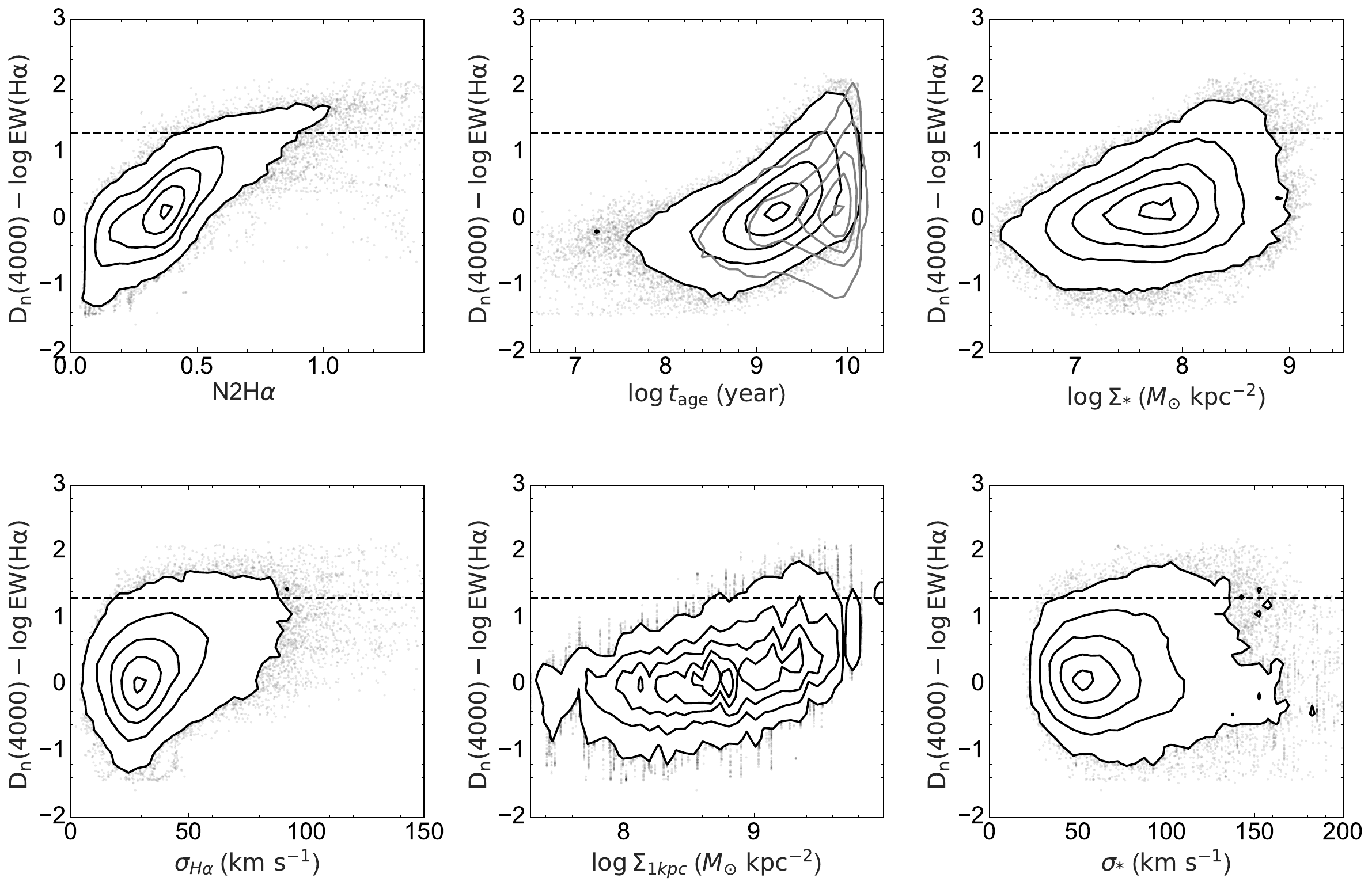}
	\caption{The correlation between \Qtracer\ and the properties with top 6 highest Gini importance. The dashed horizontal line shows the criterion we used to select quenched region in \autoref{fig:GRQRs}. The meanings of contours and black dots are same as \autoref{fig:GRQRs}. The gray contours in top center panel show the joint distribution of mass weighted age and \Qtracer. \label{fig:Q_tracer_correlation}}
\end{figure*}

\begin{figure}[ht!]
	\centering
	\includegraphics[width=0.45\textwidth]{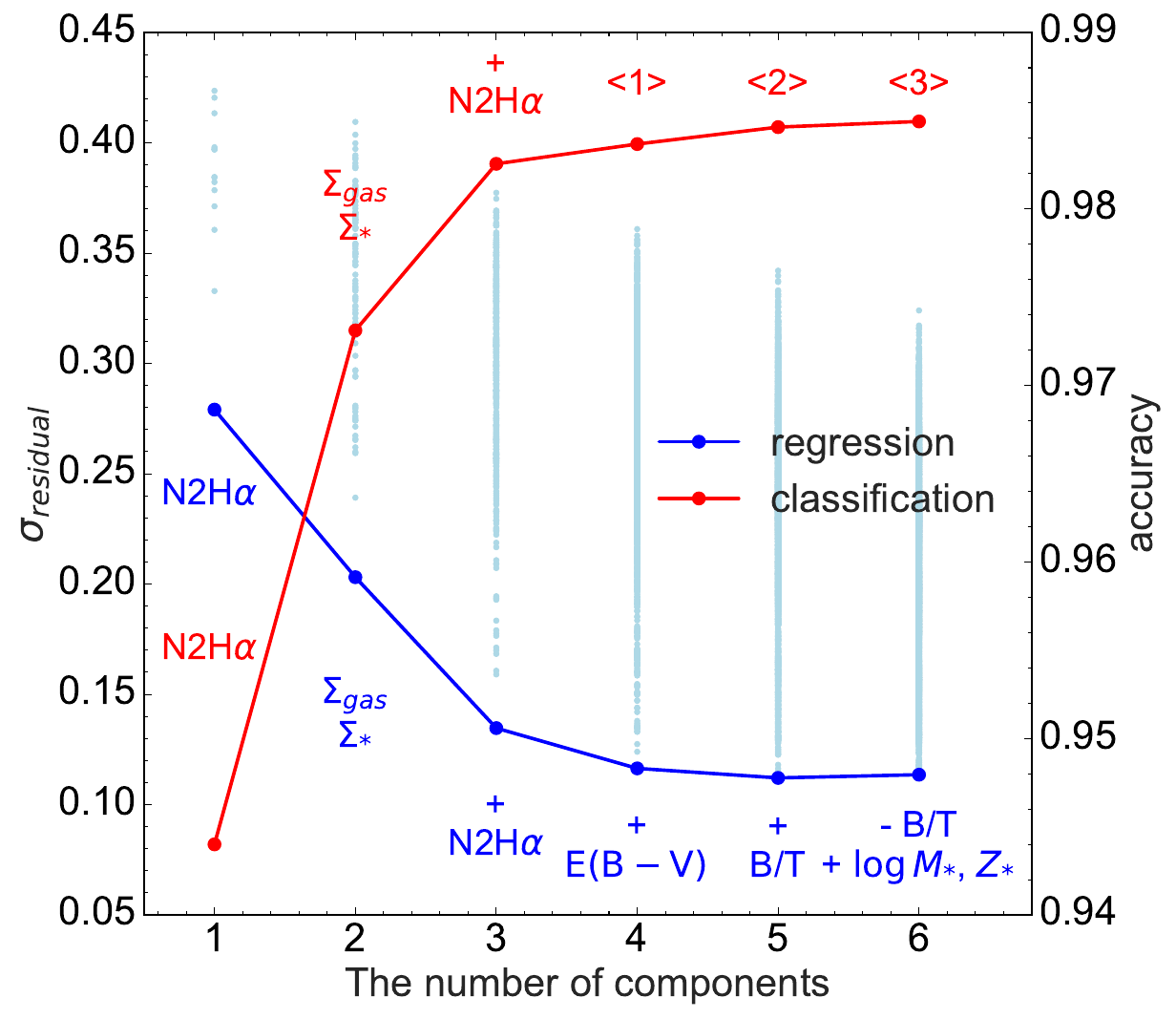}
	\caption{The figure displays the standard deviation of the residuals for the best combination with a given number of features ($N_{\mathrm{comp.}}$) in blue, while the standard deviation of residuals for other combinations are depicted as light blue dots.
	The red line demonstrates the best balanced accuracy of classifying the quenching state with a given $N_{\mathrm{comp.}}$.
	The label $\langle 1 \rangle$ denotes $\Sigma_{*}$, $\Sigma_{\rm gas}$, $\mathrm{E(B - V)}$, and $\mathrm{R/Re}$; $\langle 2 \rangle$ represents $\Sigma_{*}$, $\Sigma_{\rm gas}$, $\log M_{*}$, $\Sigma_{1\mathrm{kpc}}$, $\sigma_{\rm c}$; and $\langle 3 \rangle$ corresponds to $\Sigma_{*}$, $\Sigma_{\rm gas}$, N2H$\alpha$, $\log M_{*}$, $\Sigma_{1\mathrm{kpc}}$, $\sigma_{\rm c}$.
	% The center panel presents the results of regression for the best combination when $N_{\mathrm{comp.}} = 3$, with $\mathrm{E(B-V)}$ colored. The feature $\mathrm{E(B-V)}$ is included when $N_{\mathrm{comp.}} = 4$. In the right panel, the regression results for the best combination when $N_{\mathrm{comp.}} = 4$ are depicted, with $B/T$ colored. The feature $B/T$ is newly added when $N_{\mathrm{comp.}} = 5$.
	\label{fig:RF_combination}}
\end{figure}

\begin{figure*}[ht!]
	\centering
	\includegraphics[width=\textwidth]{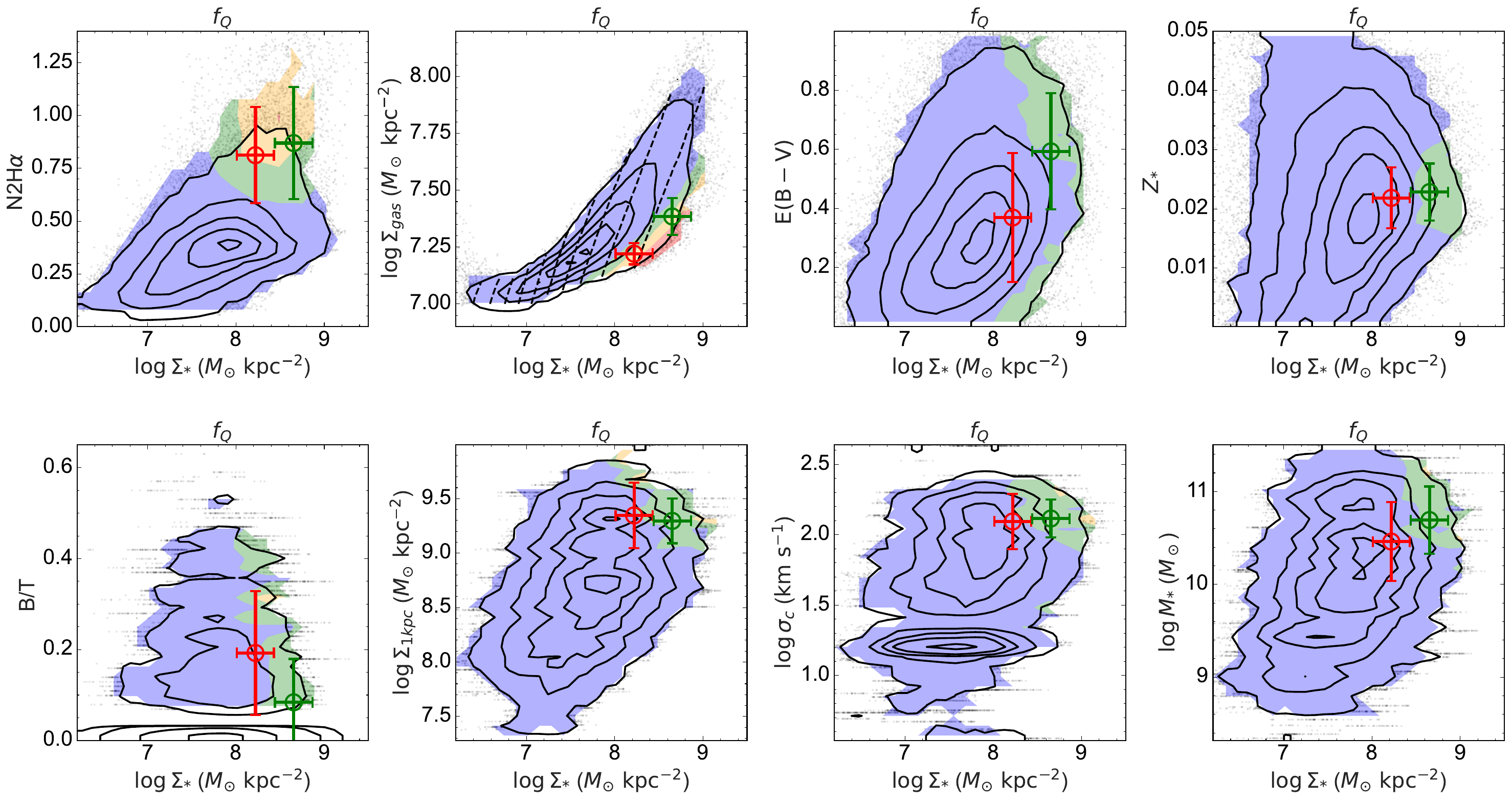}
	\caption{Each panel is a y-axis versus $\log \Sigma_{*}$ diagram. The colors blue, green, yellow, and red correspond to the \fq\ intervals of 0.0 - 0.1, 0.1 - 0.5, 0.5 - 0.9, and 0.9 - 1.0 separately. The y-axes in the first row represent local parameters N2H$\alpha$, $\log \Sigma_{gas}$, $\mathrm{E(B-V)}$, and $Z_{\ast}$, while the y-axes in the second row denote global parameters $B/T$, $\log \Sigma_{\rm 1kpc}$, $\log \sigma_c$, and $\log M_{\ast}$. The meanings of black contours and dots are same as \autoref{fig:GRQRs}. The mean value and 1 $\sigma$ scatter of GRQRs and GPQRs are shown as red and green circles with error bar, separately. In the second panel of the first row, the contours of $f_{\mathrm{gas}}$ are shown as black dashed lines. \label{fig:Q_tracer_correlation_2d}}
\end{figure*}

In this section, we investigate the significance of resolved properties in 
relation to the quenching of the selected spaxels. We have a range of 
resolved properties available, and there may be potential interaction 
effects among these properties. Therefore, we employ the Gini importance 
value provided by the random forest classifier as a data-driven method 
to identify the important properties while considering the interaction effects. 
The random forest algorithm is implemented using Scikit-learn 
\citep{scikit-learn}. Although our focus is on resolved properties in this 
section, we also include some global properties in the input of the 
random forest classifier for completeness. The resolved properties include various parameters that carry local stellar population information. These include $\Sigma_{\ast}$ (stellar surface mass density), $Z_{\ast}$ (stellar metallicity), $\log t_{\rm age}$ (logarithm of stellar age), and $\Sigma_{\rm SFR}$ (star formation rate surface density). Additionally, the analysis incorporates kinematic information such as $\sigma_{\rm H\alpha}$ (H$\alpha$ gas velocity dispersion) and $\sigma_{\ast}$ (stellar velocity dispersion). Gas phase information is also considered, including $\mathrm{E(B-V)}$ (gas phase color excess), $\Sigma_{\rm gas}$ (gas surface mass density), N2H$\alpha\equiv$ \nii/\ha\ (ratio of \nii\ to \ha\ flux), and O3N2 $\equiv$ $\log$ (\oiii/\hb) - $\log$ (\nii/\ha). Finally, the location within the host galaxy is represented by $\mathrm{R/Re}$, which denotes the galactic-centric radius normalized by the effective radius.
The global properties include 
$\log M_{\ast}$ (logarithm of global stellar mass), 
$B/T$ (bulge-to-total luminosity ratio), 
$\Sigma_{\rm 1kpc}$ (stellar surface mass density within central 1 kpc), 
and $\sigma_c$ (central stellar velocity dispersion). 
% $\Sigma_{\ast}^{0.5} \Sigma_{\mathrm{gas}}$ (product of the square root 
% of stellar surface density and gas surface density),
% $f_{\mathrm{gas}} = \Sigma_{\rm gas}/(\Sigma_{\rm gas} + \Sigma_{\ast})$ 
% (gas fraction)
For stellar age, we utilize the luminosity-weighted version 
as it is more strongly correlated with the quenching process compared 
to the mass-weighted age. This result aligns with the well-known correlation between \haew, \dnbreak, and luminosity-weighted stellar age \citep[see][for recent reviews]{Sanchez2020, Sanchez2021_global_to_resolved, Sanchez2022_MaNGA_Pipe3D, Sanchez2023_eCALIFA}. If we were to use both luminosity-weighted 
and mass-weighted age, these two properties would share their importance 
with each other, leading to an underestimation of their individual contributions. 
Similarly, for stellar metallicity, we employ the luminosity-weighted version 
for self-consistency. We note that, although SFE (defined as $\Sigma_{\rm SFR} / \Sigma_{\rm gas}$) 
	and cold gas fraction $f_{\rm gas}$ (defined as $\Sigma_{\rm gas} / (\Sigma_{\rm gas} + \Sigma_{\ast})$) are expected play important roles in understanding star formation quenching, we have chosen not to 
	include them in the random forest analysis for the following two considerations. 
	First, both SFE and $f_{\rm gas}$ are mathtically related to SFR, $\Sigma_{\rm gas}$ and $\Sigma_{\ast}$, 
	which are already included in our analysis. Inclusion of mathtically-related parameters 
	would create redundancy, and thus artificially diminish the importance of these parameters, known as the "dilution effect". In this case, we prefer to include original obseverables and more-directly derived parameters, rather than less-directly derived parameters. Secondly and more importantly, it would be difficult to interpret 
	the importance of SFE and $f_{\rm gas}$ individually, as each is a combination of two underlying parameters. For instance, a lower gas mass at a given stellar mass or a higher stellar mass at a given gas mass could both lead to a low $f_{\rm gas}$.

To address the imbalance between non-quenched and quenched spaxels, 
we randomly select a subsample of non-quenched spaxels that matches 
the number of quenched spaxels. This subsample is used alongside the 
quenched spaxels for training the random forest classifier. The training/test 
set is randomly split in an 8:2 ratio. The classifier achieves an accuracy of 
approximately 98\% on the test set, indicating that the considered physical 
properties as a whole provide complete description of the quenching status
of resolved regions. We estimate the uncertainty of importance 
by re-selecting the subsample of non-quenched spaxels and re-splitting 
the training/test set. The importance of the properties and the uncertainty 
are depicted in \autoref{fig:RF_importance}, where the properties are ordered 
by decreasing the importance. As shown, the most important property is 
N2H$\alpha$, followed by $\log t_{\rm age}$,
$\Sigma_{\ast}$, $\sigma_{\rm H\alpha}$, $\Sigma_{\rm 1kpc}$ and $\sigma_{\ast}$. The order 
of importance indicates that the quenching of local regions is primarily 
related to local physical properties rather than global properties. Importantly, it should be noted that the feature importance attributed by the random forest method only denotes correlation and not necessarily causation. In addition, to test the potential bias in the importance ranking caused by the estimated \Siggas, we have repeated the analysis by excluding \Siggas\  and found no significant changes in the importance ranking of other features.
%It is noteworthy that $\Sigma_{\ast}$ still exhibits high importance, despite 
%the fact that our selected quenched spaxels already represent spaxels with 
%relatively lower SFR at a given $\Sigma_{\ast}$.

To gain a perceptual understanding, we plot the correlation between our 
quenching parameter, \Qtracer\ and the top six important properties in 
\autoref{fig:Q_tracer_correlation}. The horizontal dashed line indicates the 
criterion for selecting the quenched regions. For stellar age, we additionally 
show the result for mass-weighted stellar age for comparison with the 
luminosity-weighted age.  As expected, the mass-weighted age is 
systematically higher than the luminosity-weighted age as the former is 
dominated by old populations, and for the same reason the luminosity-weighted 
age exhibits a stronger correlation with \Qtracer\ when compared to 
the mass-weighted age. As the top-ranked property, N2H$\alpha$ indeed 
exhibits an obvious correlation with \Qtracer, but this property still spans 
a large dynamical range for the quenched regions. For $\log t_{\rm age}$, 
$\Sigma_{\ast}$, and $\Sigma_{1kpc}$, the quenched regions have a narrow 
dynamical range in these properties, but not all regions with these properties 
falling in this range are necessarily quenched. On the other hand, quenched regions demonstrate relatively higher $\sigma_{H\alpha}$ and $\sigma_{\ast}$ compared to non-quenched regions, which aligns with the overall trend identified by \cite{Law21}, who found that LI(N)ER spaxels exhibit higher $\sigma_{H\alpha}$ and $\sigma_{\ast}$. However, the dynamical range for $\sigma_{H\alpha}$ and $\sigma_{\ast}$ is wide, and their correlations with \Qtracer\ are not strong. This could be attributed to our focus on the disk component of late-type galaxies. It is widely acknowledged that the spheroidal component is predominantly older and pressure-dominated. These findings, coupled with the high accuracy of the random forest classifier, suggest that quenching cannot be predicted by a single property but rather by a combination of multiple properties.

We further explore different combinations of the properties that can effectively 
predict the quenching state of the resolved regions. For this purpose we 
consider all possible parameter combinations, and for a given combination
we train the random forest regressor for predicting \Qtracer. In \autoref{fig:RF_combination}, the five-fold cross-validation residual scatters 
$\sigma_{\rm residual}$ 
for all the parameter combinations are plotted as the light-blue dots against 
the number of parameters involved in the combinations. The combinations 
with the smallest value of $\sigma_{\rm residual}$ at given parameter number 
are highlighted as the blue dots, connected by the solid blue line. 
When only one parameter is utilized, N2H$\alpha$ emerges as the most effective 
choice, consistent with the analysis above. When combining two of the properties, 
we find the most effective combination is the one formed by $\Sigma_{\ast}$ 
and $\Sigma_{\rm gas}$, which significantly lower down the residual scatter 
when compared to  N2H$\alpha$ alone or any other two-parameter combinations. 
These two parameters together with N2H$\alpha$ form the 
most effective combination of three parameters, with which $\sigma_{\rm residual}$ 
drops significantly. As we incorporate additional parameters, $\mathrm{E(B-V)}$ is introduced for four-parameter combinations, and $B/T$ for five-parameter combinations. However, the reduction in $\sigma_{\rm residual}$ is relatively limited.

% The center panel of the same figure displays the best-fit relationship between \Qtracer\ and the combination of $\Sigma_{\rm gas}$, $\Sigma_{\ast}$, and N2H$\alpha$, with the contours of $\mathrm{E(B-V)}$ overlaid as colorful lines. It is evident that the deviation of individual spaxels from the 1:1 relation is not at random but rather varies systematically with $\mathrm{E(B-V)}$, indicating a residual dependence of \Qtracer\ on this parameter. The right-hand panel depicts the best-fit result of combining $\Sigma_{\rm gas}$, $\Sigma_{\ast}$, N2H$\alpha$, and $\mathrm{E(B-V)}$, with the contours of the fifth property $B/T$ also overlaid as colorful lines. Notably, the contours of $B/T$ are almost perpendicular to the 1:1 relation, suggesting that $B/T$ plays a negligible role in indicating quenching, given that the first four properties are already taken into account.
% According to this analysis, we can find beyond the most important parameter N2H$\alpha$, the combination of $\Sigma_{\rm gas}$ and $\Sigma_{\ast}$ is also powerful for predicting the quenching state for spaxels in our sample.

In addition to the regression analysis, we also use the different combinations of the properties to 
train the random forest classifier in order to find the best combinations 
that can predict whether a region is quenched or not. The best five-fold 
cross-validation balanced accuracy as a function of the number of parameters
is plotted as the red line in the \autoref{fig:RF_combination}. 
The best combinations involving one, two, and three parameters are the same 
as those resulted from the regression analysis, as indicated in the figure. 
The classification accuracy is $>98\%$ for the three-parameter combination 
of $\Sigma_{\rm gas}$, $\Sigma_{\ast}$ and N2H$\alpha$. For combinations 
involving a larger number of parameters, we find different parameters may be 
used in cases of different parameter numbers, but $\Sigma_{\rm gas}$ and 
$\Sigma_{\ast}$ are always simultaneously included in the combinations. 
This result again implies the important roles of both $\Sigma_{\rm gas}$ 
and $\Sigma_{\ast}$ for quenching.

It is worth noting that this combination analysis is based on the estimated \Siggas, which relies on parameters used to define quenching. Although our test in \autoref{fig:residual} shows no bias in the estimated \Siggas\ with respect to \dnbreak\ and \haew\ in the first order, one may still worry about the potential risk associated with the combination of parameters, and the results of this combination analysis should be interpreted with caution. Our result may be verified or refined in future works using real observations of \Siggas\ or an estimator that does not explicitly rely on \haew.

We further investigate the dependency of quenching on the combination of 
different parameters by \autoref{fig:Q_tracer_correlation_2d}. To emphasize 
the role of $\Sigma_{\ast}$, each panel in the figure displays the distribution 
of the parent sample of spaxels on the diagram of a second property versus
$\Sigma_{\ast}$, plotted as solid black contour lines. The secondary properties 
considered are those that appear in the best combination of our random forest 
classifier or regressor. In each panel, the blue, green, yellow, and red colors
correspond to the \fq\ intervals of 0.0 - 0.1, 0.1 - 0.5, 0.5 - 0.9, and 0.9 - 1.0, respectively. Here, \fq\ represents the fraction of quenched regions meeting our quenching selection criterion. 
As can be seen, fully quenched regions with $f_{Q}>0.9$ can be selected out 
only through a combination of \Siggas\ and \Sigstar\ (the second panel in the upper row). 
Additionally, regions with $f_{Q}>0.5$ can be identifiable by combining 
\Sigstar\ with N2H$\alpha$, as shown in the leftmost panel in the upper row.
In other panels there is nowhere for $f_{Q}\ge 0.5$, indicating that none of the 
pairwise combinations of those properties are sufficient for predicting quenching.
In each panel the mean and 1-$\sigma$ scatter of GRQRs and GPQRs are 
additionally plotted as the green and red circles with error bars. Only in the 
first two panels in the upper row, as expected, the locations of the quenched regions agree 
with the quenched regions indicated by the contours of high $f_{Q}$. 
This result shows that the combinations of $\Sigma_{\ast}$ with either 
N2H$\alpha$ or $\Sigma_{\rm gas}$ provide sufficient criteria for quenching. We have repeated the above analysis for all the other pairwise 
combinations using the properties that appear in this plot, and found none of 
them can indicate quenching as efficiently as the combinations in the first two 
panels of \autoref{fig:Q_tracer_correlation_2d}. 
% Comparing the first two panels, we find the combination of $\Sigma_{\ast}$
% with $\Sigma_{\rm gas}$ in the second panel to be more efficient. 
% It is natural to expect low gas densities
% for quenched regions, but the importance of high stellar densities is not 
% naturally expected from our common knowledge.

In the panel of $\Sigma_{\ast}$ versus $\Sigma_{\rm gas}$, we additionally show the distribution of the gas fraction $f_{\mathrm{gas}}$ as dashed black contours. It is evident that the overall trend of $f_{\mathrm{gas}}$ aligns with the contour lines of $f_{Q}$, emphasizing the significant role of $f_{\mathrm{gas}}$ in quenching. As aforementioned, however, attributing the lower $f_{\mathrm{gas}}$ solely to a lower gas mass at a given stellar mass can only account for the drop from the rSFMS, but not the drop from the rSK law which requires a decrease in the SFE at a given gas mass. Otherwise, the quenched region would only shift along the rSK law from the upper-right to the lower-left. Therefore, our result emphasizes the importance of the combination of $f_{\mathrm{gas}}$ and SFE for quenching, which is broadly consistent with previous studies  \citep{Colombo2018, Colombo2020, Ellison2020, Ellison2021_nonuni, Sanchez2021, Kalinova2022}. This appears to conflict with the finding that the combination of \Siggas\ and \Sigstar\ can efficiently select quenched regions, as the SFE is not immediately known with the combination of these two parameters. A natural explanation is that the lower $f_{\mathrm{gas}}$ should also be considered as the presence of more stars at a given gas mass, which will have a net negative effect on star formation, leading to the decrease of SFE.

Further insight can be gained by repeating our random forest analysis for the GRQRs and GPQRs  separately. The results are shown in  \autoref{fig:RF_importance}. For GRQRs, the importance of \Sigstar\ and \sigstar\ is significantly enhanced, while the importance of \Sigsfr\ is suppressed. In contrast, for GPQRs, the importance of \Sigsfr\ is enhanced to be comparable to that of \Sigstar. This result suggests that, for GRQRs, the decrease of SFE at a given \Siggas\ is highly correlated with the high \Sigstar, which leads to a lower $f_{\mathrm{gas}}$ at the same time. For GPQRs, the decrease of SFE is much more independent of the value of \Sigstar, thus the additional information of \Sigsfr\ should be included to effectively select such regions. The enhancement of the importance of \Siggas\ is trivial, as GRQRs and GPQRs are separated by \Siggas. The enhancement of the importance of $\mathrm{R/Re}$ is influenced by the \Sigstar\ profile of the galaxy, as GRQRs tend to be found in the inner disk, which have a higher \Sigstar.

In conclusion, we identify N2H$\alpha$, $\Sigma_{\rm gas}$, and $\Sigma_{*}$ as the most significant properties associated with quenching. For gas-rich regions, $\Sigma_{\ast}$ is a particularly significant property, which potentially drives the simultaneous decrease of $f_{\mathrm{gas}}$ and SFE. For GPQRs, the decrease of SFE is also necessary but is more independent of \Sigstar. Additionally, GRQRs exhibit higher $\mathrm{E(B-V)}$ than GPQRs. These differences should be attributed to the distinct quenching mechanisms associated with the different gas content of GRQRs and GPQRs, a topic that will be discussed in \autoref{sec:discussion}.

\begin{figure*}[ht!]
	\centering
	\includegraphics[width=\textwidth]{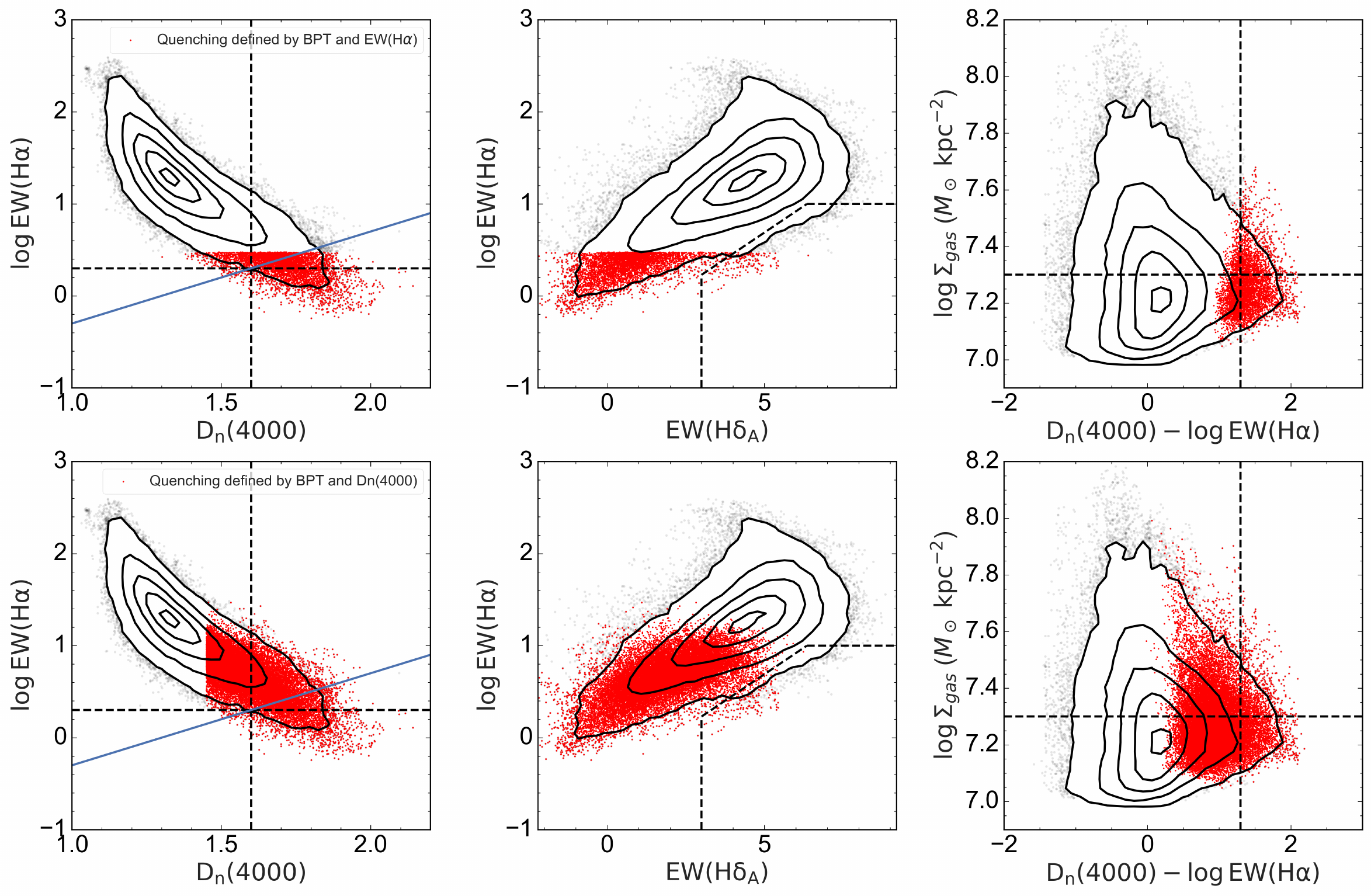}
	\caption{The distribution of resolved regions of our sample on the diagram
		of $\log$\haew\ versus \dnbreak\ (left), $\log$\haew\ versus 
		\hdew\ (center), and $\log$\Siggas\ versus \Qtracer\ (right). 
		The red dots in each panel represent the subsample of ``quenched regions'', which 
		are defined jointly by BPT diagrams and \haew$<3$\AA\ following 
		\cite{Lin2019-quenching} (upper panels), or jointly by BPT diagrams and 
		\dnbreak$>1.45$ following \cite{Bluck2020-global-local,Bluck2020-central-satellite} (lower panels).
		\label{fig:quenching_definition}}
\end{figure*}

\begin{figure*}[ht!]
	\centering
	\includegraphics[width=\textwidth]{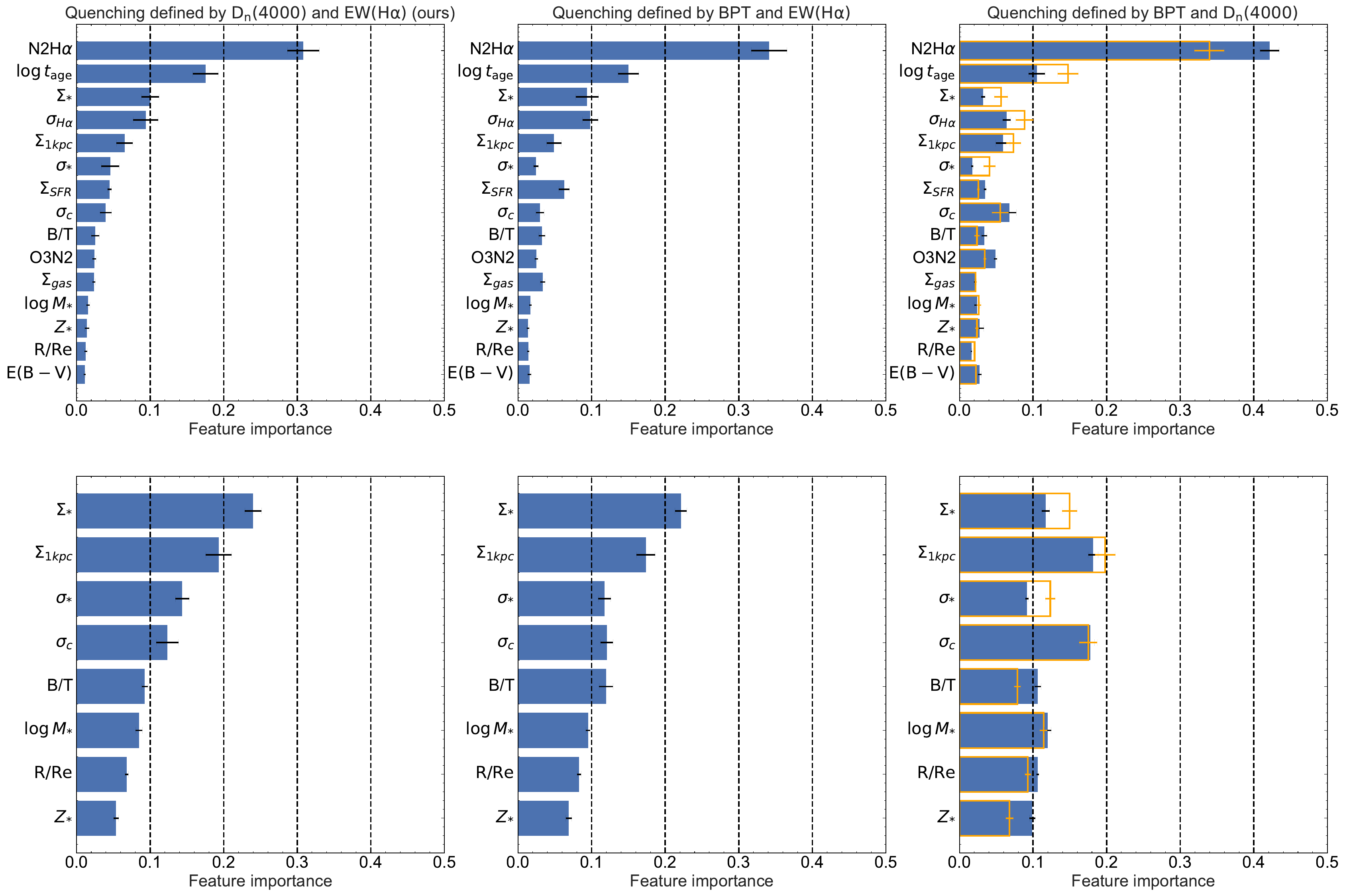}
	\caption{Gini feature importance provided by the random forest classifier for 
		quenched regions defined in three different ways (panels from left to right):
		by \dnbreak\ and \haew\ in this work, by BPT diagrams and \haew\ following 
		\cite{Lin2019-quenching}, and by BPT diagrams and \dnbreak\ following 
		\cite{Bluck2020-global-local,Bluck2020-central-satellite}. In the upper panels, the same set of properties as used 
		in \autoref{fig:RF_importance} are included in the analysis here. For the lower panels, 
		a subset of properties similar to those used in \cite{Bluck2020-global-local,Bluck2020-central-satellite} are included
		in the analysis. The properties are ordered by decreasing the feature importance 
		as resulted from the samples in the leftmost panel. The orange empty histograms
		in the rightmost panels show the results when the criterion of \dnbreak\ $>1.6$ 
		is adopted instead of \dnbreak\ $>1.45$.
		\label{fig:rf_quenching_definition}}
\end{figure*}

\subsection{Dependence on quenching definition}\label{sec:diff_definition}

The quenched regions in our sample are selected to simultaneously have 
\haew$<2$\AA\ (thus with no/weak ongoing star formation) and \dnbreak$>1.6$
(thus with no/little stellar populations younger than 1-2 Gyr). This definition 
has been widely adopted in previous studies \citep[e.g.][]{Geha2012,2018ApJ...856..137W}. 
For simplicity, we have combined the two criteria and introduced a single-parameter 
selection: \Qtracer$\ge 1.3$, which is shown to result in a sample similar to the one 
selected jointly by the two separate criteria (see \autoref{fig:GRQRs}). On the 
other hand, as mentioned, some other studies have used different definitions 
for quenching. 
For instance, \citet{Lin2019-quenching} defined quenched regions as non-star-forming 
regions on BPT diagrams that have \haew$<3$\AA, while \citet{Bluck2020-global-local,Bluck2020-central-satellite} 
also used BPT diagrams to select non-SF regions and essentially 
used \dnbreak\ to further select regions of low sSFR. 
In order to examine the dependence on quenching definition, we have selected 
two new samples of ``quenched regions'' from our parent sample, by applying 
the definitions in \cite{Lin2019-quenching} and \cite{Bluck2020-global-local,Bluck2020-central-satellite}. We note that,
for the definition of \cite{Bluck2020-global-local,Bluck2020-central-satellite}, we didn't attempt to estimate sSFRs by 
\dnbreak\ for passive regions and then select quenched regions by sSFRs. 
Instead, we simply apply a cut of \dnbreak$>1.45$ to the non-star-forming regions,
which results in closely matching samples of quenched regions as discussed
in \cite{Bluck2020-global-local,Bluck2020-central-satellite}. 

In \autoref{fig:quenching_definition}, we show the distribution of the two samples 
of quenched regions in the diagrams formed by \haew, \dnbreak, \hdew, \Siggas\ 
and \Qtracer, in the same way as in \autoref{fig:GRQRs}.
When compared to the quenched regions in our sample, the quenched regions 
defined jointly by \haew\ and BPT diagrams as in \cite{Lin2019-quenching} cover 
similar ranges in \dnbreak, \hdew\ and \Qtracer, although they have slightly stronger 
H$\alpha$ emission due to the higher \haew\ threshold.  
In contrast, the quenched 
regions defined by \dnbreak\ and BPT diagrams as in \cite{Bluck2020-global-local,Bluck2020-central-satellite} cover wider 
ranges in those parameters, extending to much higher \haew\ and lower \dnbreak. 
The similarity of our sample to \cite{Lin2019-quenching} can be understood 
from the similar cuts in \haew\ as adopted in both studies. As can be seen from 
the \haew-\dnbreak\ diagram, regions with substantially low \haew\ appear to 
mostly have the highest values of \dnbreak, while a substantially high \dnbreak\ 
doesn't necessarily lead to low \haew. This fact also explains the difference of
our sample from that of \cite{Bluck2020-global-local,Bluck2020-central-satellite}, which is selected by \dnbreak\ but 
not \haew. Therefore, the differences in the different samples of quenched regions
are essentially caused by the different requirements on H$\alpha$ emission.

We then apply the random forest classifier to identify important properties related to quenching in both samples. We first consider the same set of properties as 
analyzed above for our sample. The results are shown in the upper panels in 
\autoref{fig:rf_quenching_definition}. For comparison, the result of our sample 
is repeated here as the upper left panel. We see that N2H$\alpha$ is top ranked 
in all the samples. This result is not unexpected for the two new samples, 
which are both selected by BPT diagrams involving N2H$\alpha$. We have 
repeated the analysis without N2H$\alpha$, finding the ranking of the rest 
properties remain unchanged although the absolute value of feature 
importance increases to varying degrees for all the properties. As expected, 
the sample defined following \cite{Lin2019-quenching} behaves similarly to 
our sample, with $\log t_{\rm age}$, $\sigma_{\rm H\alpha}$, 
$\Sigma_{\ast}$ and $\Sigma_{\rm SFR}$ being the most important properties. 
Differently, for the sample defined following \cite{Bluck2020-global-local,Bluck2020-central-satellite}, the most important 
properties are $\log t_{\rm age}$, $\sigma_{c}$, 
$\sigma_{\rm H\alpha}$ and $\Sigma_{\rm 1kpc}$. One may wonder whether 
the difference is caused by the lower \dnbreak\ threshold (\dnbreak$>1.45$)
adopted by \cite{Bluck2020-global-local,Bluck2020-central-satellite}. To test this out, we have increased the cut to 
\dnbreak$>1.6$ to mimic our selection and repeated the analysis. Plotted 
as the yellow histograms in the same panel, the result becomes more 
similar to our sample in the sense that the feature importance increases for 
$\log t_{\rm age}$ and $\sigma_{\rm H\alpha}$, but the relatively high 
importance of $\sigma_{c}$ and $\Sigma_{\rm 1kpc}$ and low importance 
of $\Sigma_{\ast}$ remain. In the lower panel, we 
repeat the analysis for the three samples but considering a subset of properties 
similar to those considered in \cite{Bluck2020-global-local,Bluck2020-central-satellite}. Again, our sample and the one 
defined following \cite{Lin2019-quenching} present similar results as expected.
For the third sample, $\sigma_{c}$ and $\Sigma_{\rm 1kpc}$ become most 
important, consistent with \cite{Bluck2020-global-local,Bluck2020-central-satellite}. The two properties 
are still top ranked if \dnbreak$>1.6$ instead of \dnbreak$>1.45$ is 
applied in the sample selection, as shown by the yellow histograms. This result suggests that the definition of quenching can significantly impact the ranking of feature importance, leading to different conclusions. If \haew\ is not considered in the quenching definition, the quenching of resolved regions could be efficiently predicted by $\sigma_c$ or $\Sigma_{\rm 1kpc}$. This suggests that quenching is a global process, with a significant central bulge and possibly a central supermassive black hole playing a major role in the quenching process. However, when an \haew\ cut is applied, the importance of $\sigma_c$ and $\Sigma_{\rm 1kpc}$ decreases significantly, indicating that quenching is a local process.

\begin{figure*}
	\centering
	\includegraphics[width=\textwidth]{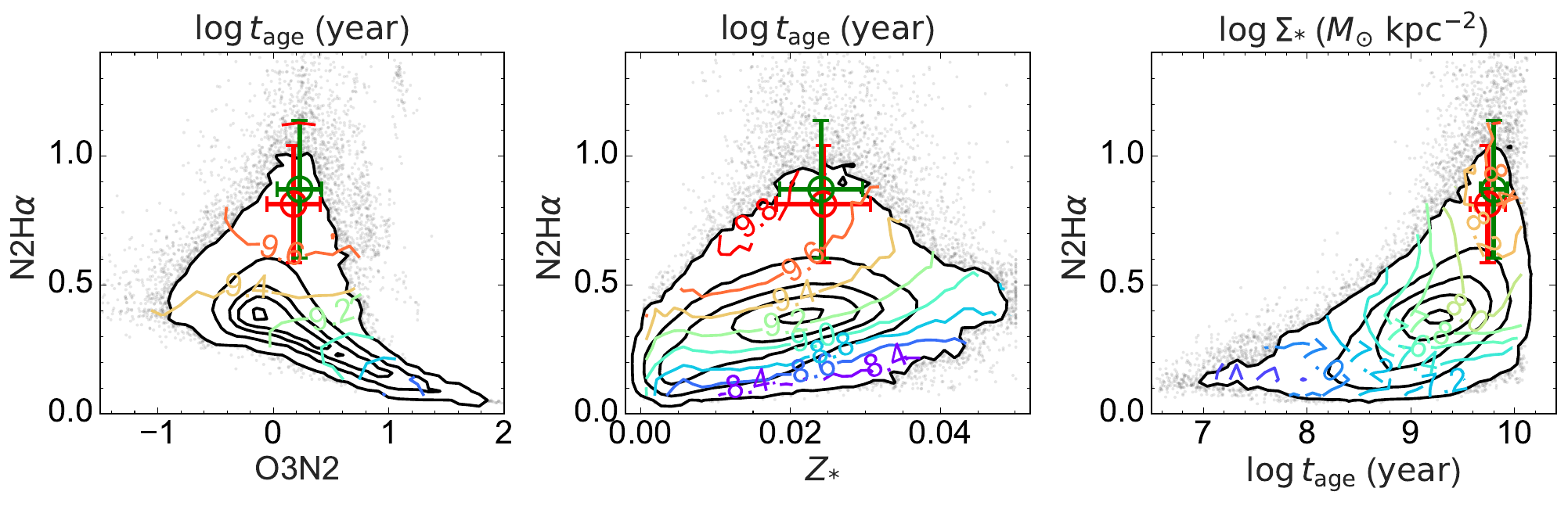}
	\caption{Each panel is a N2H$\alpha$ versus x-axis diagram. The x-axes are O3N2, $Z_*$, and $\log t_{\mathrm{age}}$ from left to right. The meanings of black contours and dots are same as \autoref{fig:GRQRs}. The mean value and 1 $\sigma$ scatter of GRQRs and GPQRs are shown as red and green circles with error bar, separately. In left and center panels, the colorful contours show the distribution of $\log$ stellar age. In right panel, the colorful contours represent the $\log \Sigma_{*}$ distribution. \label{fig:N2Ha_correlation}}
\end{figure*}

It is important to note that the difference between our sample and \cite{Bluck2020-global-local,Bluck2020-central-satellite} lies not only in the definition of quenching but also in our focus on the disk component of late-type galaxies. Our sample should be considered as a subset of the sample used by \cite{Bluck2020-global-local,Bluck2020-central-satellite}. Therefore, the disparities between our findings and those of \cite{Bluck2020-global-local,Bluck2020-central-satellite} can partly be attributed to quenching mechanisms varying in galaxies with different morphological types and in different components of late-type galaxies.

Nevertheless, within our sample, whether a \haew\ cut is used in selecting quenched regions also leads to differing results, as we have previously demonstrated. As depicted in \autoref{fig:quenching_definition}, the sample defined following \cite{Bluck2020-global-local,Bluck2020-central-satellite} covers a larger parameter space than the sample defined following \cite{Lin2019-quenching}, indicating that the samples defined with \haew\ are a subset of the sample defined by \cite{Bluck2020-global-local,Bluck2020-central-satellite}.
Therefore, the key to understanding the differences in quenching definitions lies in comprehending the physical origins of the regions with \haew $\ge$ 3 \AA\ but satisfy the quenching definition based on the combination of the BPT diagram and \dnbreak.
We propose two possible scenarios for these regions. In one scenario, these regions are intrinsically quenched, and the high \haew\ implies the presence of AGN ionization. Consequently, the selection criteria with \haew\ cut would exclude the AGN-related quenching regions, leading to a potentially underestimated importance of AGN indicators (e.g., $\sigma_{c}$ and $\Sigma_{\rm 1kpc}$). In the other scenario, both AGN and OB stars contribute to the ionization of these regions. In this case, the selection criteria without the \haew\ requirement essentially identify the correlation between AGN ionization tracer and other AGN tracers. Although the stellar population information is used to constrain the fraction of young stars (\dnbreak\ or spectral fitting-based \Sigsfr), if the star formation region is heavily obscured, the spectrum would be dominated by a less extincted, older population. Neither the spectral index like \dnbreak\ nor the typical spectrum fitting approach that uses a single attenuation curve for all star populations can efficiently capture the existence of newly formed stars. 
The hint is obtained by further subdividing the quenched regions selected based on \dnbreak\ and the BPT diagram by \haew. The subsample with \haew $\ge$ 3 \AA\ has an overall lower $\sigma_{\rm H\alpha}$ (33, 43, 55 km s$^{-1}$ for the 25\%, 50\%, 75\% quantiles) than the subsample with \haew $<$ 3 \AA\ (43, 56, 72 km s$^{-1}$). This suggests that part of $H\alpha$ emission lines from these \haew $\ge$ 3 \AA\ regions could be attributed to the gas embedded in the disk, possibly the star-forming region, as quenched regions and AGN-dominated regions tend to have higher $\sigma_{\rm H\alpha}$ \citep{Law21, Sanchez2024}.
The actual situation may involve a combination of these two scenarios or be much more complex. It is also likely that \haew-selected samples represent a later stage of the quenching process for a resolved region than the samples selected by the BPT diagram and \dnbreak\ alone. In this sense, both global quenching as traced by $\sigma_{c}$ and $\Sigma_{\rm 1kpc}$ and local quenching as indicated by N2H$\alpha$, $\Sigma_{\rm gas}$, and $\Sigma_{\ast}$ can be at work, but at different stages, during the overall quenching process of the host galaxy. 
Since our sample can be considered a subset of the sample selected without the requirement of \haew, our findings suggest that AGN feedback cannot be deemed the sole quenching mechanism.
We underscore the importance of carefully handling the regions with significant AGN features when selecting and studying quenching regions, as this can significantly impact the results. However, gaining a deeper understanding of this issue will require further efforts in the future.

\subsection{Understanding the importance of N2H$\alpha$}\label{sec:N2Ha_as_result}

The [N{\sc ii}]-to-H$\alpha$ line ratio, N2H$\alpha$, is top ranked in the 
Gini feature importance analysis regardless of quenching definition 
(\autoref{fig:RF_importance} and \autoref{fig:rf_quenching_definition}), 
and this parameter indeed shows a significant correlation with \Qtracer\ as seen 
from \autoref{fig:Q_tracer_correlation} and \autoref{fig:RF_combination}.
For the quenched regions defined by BPT diagrams but with no requirement 
on H$\alpha$ emission, the high importance of N2H$\alpha$ could be partly interpreted 
as a consequence of AGN-related ionization processes. However, for the quenched regions defined by \haew\ as in our work, an alternative explanation is necessary. In regions primarily ionized by OB stars, N2H$\alpha$ can serve as an indicator of gas phase metallicity (e.g. \citealt{2004MNRAS.348L..59P}). Nevertheless, for quenched spaxels typically found in the LI(N)ER region of the BPT diagram and manifesting weak \ha\ emission, the contribution of ionizing photons from low-mass evolved stars cannot be ignored \citep{Binette1994,Stasinska2008,Sarzi2010,2012ApJ...747...61Y,2013A&A...555L...1P,2013A&A...558A..43S,Sanchez2014,2016MNRAS.461.3111B,2016A&A...588A..68G,2017MNRAS.466.2570B,Morisset2016, 2017MNRAS.472.4382R, 2017MNRAS.466.3217Z}. In this case, higher N2H$\alpha$ values indicate characteristics of the ionization source but not the metallicity. In the upper-left panel of \autoref{fig:N2Ha_correlation}, we compare N2H$\alpha$ with other widely used gas-phase metallicity tracer O3N2. The majority of spaxels in the parent sample, enclosed by the 90\% contour line, exhibit a narrow and tight relationship on this diagram, reflecting the predominance of star-forming regions in our sample. However, the quenched regions deviate from this relationship and extend to the highest N2H$\alpha$ region, corresponding to the spaxels with the highest average stellar age, as indicated by the colored contours. This demonstrates that, in the regions dominated by old population, N2H$\alpha$ and O3N2 no longer reliably represent gas-phase metallicity and lose their tight correlation. The center panel shows the correlation of N2H$\alpha$ with stellar metallicity. Similarly, the quenched regions are found to have intermediate metallicities, located outside the 90\% contour line of the parent sample. The rightmost panel in the same figure displays the N2H$\alpha$ versus 
$\log t_{\rm age}$ diagram, which shows again that the regions with highest 
N2H$\alpha$ have the oldest stellar populations. On the other hand, however, 
the regions of the highest stellar age span a full range of N2H$\alpha$. Thus,  it is 
not necessary for regions with the highest stellar age to also have the highest 
N2H$\alpha$ values, unless they have a relatively high value of $\Sigma_{\ast}$ 
as shown by the colored contour lines. This result may be related to the fact 
that low mass evolved old stars are not as bright as massive main sequence stars. Given that the mass is primarily contributed by low-mass stars, a higher stellar mass density is needed to enable evolved stars to become the dominant ionizing source. 
The correlations between N2H$\alpha$, $t_{\rm age}$, $\sigma_{H\alpha}$, \sigstar\ and the parameters we used for selecting quenching regions are well established in previous studies \citep{Binette1994,Stasinska2008,Sarzi2010,2012ApJ...747...61Y,2013A&A...555L...1P,2013A&A...558A..43S,Sanchez2014,2016MNRAS.461.3111B,2016A&A...588A..68G,2017MNRAS.466.2570B,2017MNRAS.466.3217Z, Sanchez2020, Sanchez2021_global_to_resolved, Law21, Sanchez2022_MaNGA_Pipe3D, Sanchez2023_eCALIFA}. The quenched regions tend to have higher values of N2H$\alpha$, $t_{\rm age}$,  $\sigma_{H\alpha}$, \sigstar\ and \dnbreak, as well as lower \haew. The importance of N2H$\alpha$, $t_{\rm age}$, $\sigma_{H\alpha}$, and \sigstar\ to quenching can be understood as a result of much more diffuse gas ionization by photons from substantially large amounts of low-mass evolved stars that typically have higher velocity dispersion. In this regard, the importance of these four parameters is correlated with each other and is a natural consequence rather than a cause of the quenching process.

\section{Discussion}\label{sec:discussion}

\subsection{Implications of the importance of $\Sigma_{\ast}$}

The most striking result from our work is the high importance of $\Sigma_{\ast}$ 
to quenching. This property appears to be not important at all as a single indicator.
However, as illustrated in \autoref{fig:Q_tracer_correlation_2d}, the combination of low $\Sigma_{\rm gas}$ and high $\Sigma_{\ast}$ is not only sufficient but also necessary for indicating quenching. The regions identified by these two features are predominantly quenched, and quenched regions typically demonstrate the unique distribution of these two features. As discussed in \autoref{sec:resolved_properties}, the high $\Sigma_{\ast}$ could result in lower $f_{\mathrm{gas}}$ and lower SFE simultaneously, making it the most important property for quenching, apart from the features that are considered as the outcome of quenching (as discussed in \autoref{sec:N2Ha_as_result}), especially for GRQRs.

Previous studies have investigated the role of $\Sigma_\ast$ on star formation 
instead of quenching, by examining the correlation of star formation rate (SFR) 
surface density $\Sigma_{\rm SFR}$ with the combination of $\Sigma_{\rm gas}$ 
and $\Sigma_{\ast}$. Both positive and negative effects have been found. 
For instance, \citet{2011ApJ...733...87S,2018ApJ...853..149S} found a power-law 
relation of $\Sigma_{\rm SFR}\propto (\Sigma_{\ast}^{0.5} \Sigma_{\rm gas})^{1.09}$, 
suggesting a positive role of $\Sigma_{\ast}$  which may promote star formation by
providing an additional local gravitational potential to help gas infall and condensation. 
Based on the ALMaQUEST sample with both integral field spectroscopy from MaNGA 
and CO intensity mapping from ALMA, \citet{Lin2019} found $\Sigma_\ast$ 
to present a negative effect on star formation, with a best-fit relation of 
$\Sigma_{\rm SFR}\propto (\Sigma_{\ast}^{-0.3}\Sigma_{\rm H_2})^{1.38}$. 
We have performed the same analysis using our sample. 
Similar to \cite{Lin2019}, our star-forming subsample reveals a negative power exponent 
of -0.31 for $\Sigma_\ast$. The discrepancy between the MaNGA-based samples 
in \cite{Lin2019} and our work and those in \citet{2011ApJ...733...87S,2018ApJ...853..149S} 
may be due to the different stellar mass ranges covered by the different samples, as 
pointed out by \citet{Lin2019}. It is important to note that, the relation 
proposed by \citet{2011ApJ...733...87S,2018ApJ...853..149S} outperforms the 
canonical Kennicutt-Schmidt relation primarily in the outer regions of dwarf galaxies 
characterized by low surface mass density or extremely metal-poor ``gas-dominated'' 
regions with gas fraction $>$0.5 (see Figure 2, 3, 4 in \citealt{2018ApJ...853..149S}). 
Conversely, our sample consists mainly of ``star-dominated'' regions with gas fraction 
$<$0.5. These findings suggest that the net effect of existing stars for star formation 
may vary according to local properties, potentially being positive in low-mass, 
metal-poor, or gas-dominated regions, but negative in regions similar to those in
this work and \cite{Lin2019}. 
\cite{Pessa2022} found that the power exponent of $\Sigma_\ast$ can vary significantly across different local environments (not to be confused with the galaxy environment) and when defined at different scales. On a scale of approximately 100 pc, the power exponent is positive for regions in the disk, spiral arms, and the center of the galaxy, while it is negative for regions in the galaxy ring and bar. Using all spaxels from different environments to calibrate the correlation results in the power exponent being nearly zero and slightly negative.
However, when the correlation is studied at a scale of 1000 pc, the power exponent for regions in different environments is similar. The power exponent defined by all spaxels will be close to -0.3, as found by this work and \cite{Lin2019}. This result implies that the effect of $\Sigma_\ast$ is complex and could differ in different local environments and at different scales. It is important to note that the role of $\Sigma_\ast$ in star-forming regions may not necessarily be the same as in quenched regions. As will be demonstrated below, the net effect of $\Sigma_\ast$ is the result of a combination of multiple competing mechanisms. Therefore, the statistical net effect may vary with the physical properties of the region, the local environment, and the spatial scale.

\begin{figure*}
	\centering
		\includegraphics[width=\textwidth]{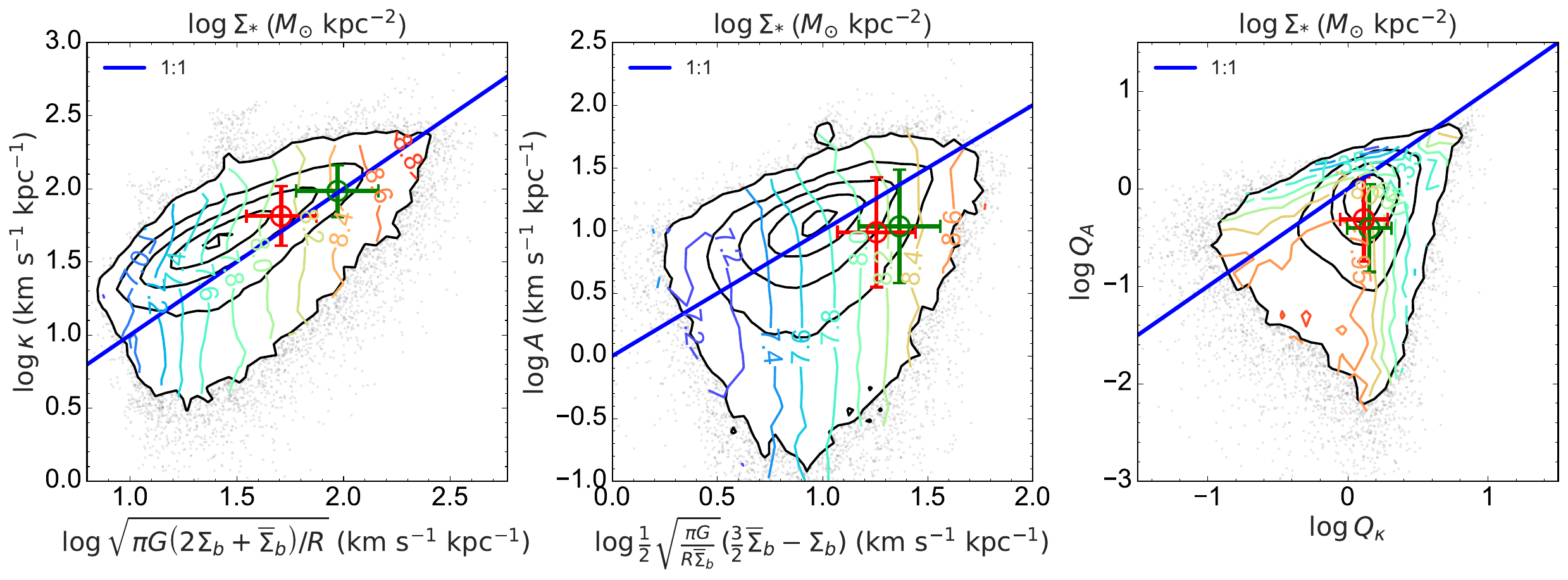}
	\caption{Left and center panels, the comparisons of estimates of $\kappa$ (left panel) and $A$ (center panel)
	with the values given by the relations in \autoref{eqn:kappa_relation} and \autoref{eqn:A_relation}. For the x-axes of these two panels, we use  $\Sigma_{\rm b}=\Sigma_{\rm gas}+\Sigma_\ast$ instead of $\Sigma_{\rm m}$. The right panel illustrates the correlation between the $Q$ value measured by $\kappa$ and $A$. The 1:1 relationship is indicated by the blue line in each panel. \label{fig:kappa}}
\end{figure*}

\subsection{Roles of existing evolved stars in quenching}

As discussed above, the importance of \Sigstar\ implies that existing 
low-mass stars play some positive roles for quenching, particularly for 
quenched regions with relatively large amounts of cold gas. 
In fact, existing stars can exert influence on surrounding gas in a number 
of different ways. The net effect on quenching is likely a combined result of 
different mechanisms. In this subsection we discuss on possible 
quenching mechanisms as driven by existing evolved stars.

\subsubsection{Dynamical stabilization}\label{sec:dynamics}

In a thin, differentially rotating disk as usually assumed in models of star 
formation, the potential for gas clouds to collapse is set 
by the competition between the self-gravity of the disc and the combined 
effect of thermal pressure, Coriolis forces and shear
\citep[e.g.][]{1960AnAp...23..979S,1964ApJ...139.1217T,1984ApJ...276..127J,1989ApJ...344..685K,1992MNRAS.256..307R,1994ApJ...435L.121E,1994ApJ...427..759W,1998ApJ...493..595H,2000ApJ...536..173T,2011ApJ...737...10E}.
The ability of a two-component disc with both gas and stars to stabilize 
itself through thermal pressure and Coriolis forces can be quantified by 
the modified version of the \citet{1964ApJ...139.1217T} $Q$ parameter 
as \citep{1994ApJ...427..759W}:
\begin{equation}\label{eqn:Q}
	Q=\left(\frac{1}{Q_{\mathrm{gas} }}+\frac{1}{Q_{\ast}}\right)^{-1}.
\end{equation}
Here, $Q_{\mathrm{gas}}$ and $Q_{\ast}$ are the stability parameters 
for the gas and stellar components respectively, defined as
\begin{equation}\label{eqn:Qgasstar}
	Q_{\mathrm{gas}}=\frac{\kappa c_{\rm s}}{\pi G \Sigma_{\mathrm{gas}}};~~~
%\end{equation}
%\begin{equation}
	Q_{\ast}=\frac{\kappa \sigma_{\rm r,\ast}}{\pi G \Sigma_{\ast}},
\end{equation}
where $\kappa$ is the epicyclic frequency, $c_{\rm s}$ is the characteristic 
sound speed of $10^4$ K gas, and $\sigma_{\rm r,\ast}$ 
is the radial stellar velocity dispersion. In cases where the growth of 
gas clouds occurs with streaming motions along interstellar magnetic 
field lines or inner regions of galaxies with rising rotation curves, 
the Coriolis force can be less important and the formation of dense 
clouds may involve more of a competition with shear than with 
Coriolis forces, as proposed by \citet{1998ApJ...493..595H}. In this case,
\citet{1998ApJ...493..595H} estimated that the critical surface density 
for a perturbation to grow by a sufficiently large factor of $\sim100$ is 
$\Sigma_{\rm c,A}=(2.5Ac_{\rm s})/(\pi G)$, where the Oort constant 
$A=0.5(v/R-dv/dR)$ quantifies the local shear rate at galactic-centric radius $R$. 
Accordingly, the Q parameters defined above can be rewritten by simply 
replacing $\kappa$ by $2.5A$.

Therefore, on one hand, the existing stars can facilitate star formation 
through increasing \Sigstar, hence decreasing $Q_\ast$.  On the other hand, 
these stars may contribute 
to the competition with self-gravity to locally stabilize the disk against 
fragmentation, by influencing thermal pressure, Coriolis forces and shear 
as traced by $\sigma_{\rm r,\ast}$, $\kappa$ and $A$. For $\sigma_{\rm r,\ast}$,
the positive role of existing stars is evident given the positive correlation 
between stellar velocity dispersion and surface mass density \citep[Eqn.6]{2018MNRAS.474.2323S}. However, 
the roles of existing stars for $\kappa$ and $A$ are not immediately clear. 
%As a quantity for
%Coriolis forces which come from angular momentum conservation, 
%the epicyclic frequency $\kappa$ 
%is the oscillation frequency when the 
%closed orbit is perturbed, a measure of the ability to recover from perturbation and thus the stability of a rotating disk. 
Given the rotation 
curve $v(R)$, i.e. the rotation velocity as a function of the galactic-centric 
radius $R$, one can obtain $\kappa$ at given $R$ by the definition:
\begin{equation}\label{eqn:kappa0}
	\kappa(R) \equiv \sqrt{\frac{2v}{R} \frac{d v}{d R} + \frac{2 v^2}{R^2}}.
\end{equation}
For a simple case with circular rotation, $v(R)$ is simply 
scaled with the total mass within $R$ as $v(R)\propto \sqrt{M(R)/R}$, and thus 
with the average surface mass density within $R$ as 
$v(R)\propto \sqrt{R\overline{\Sigma}_{\rm m}(R)}$, given the relation of 
$M(R) = \pi R^2 \overline{\Sigma}_{\rm m}(R)$.
One can easily find that the definition of $\kappa$ is reduced to the following relation:
%\begin{equation}
%	\kappa(r) \propto \left[\frac{\pi}{r}\left(2\Sigma_{\rm m}(r) + \overline{\Sigma}_{\rm m}(r)\right)\right]^{1/2}.
%\end{equation}
%\begin{eqnarray*}
%	\kappa(r) & \propto & \left[\frac{1}{r}\left(2\Sigma_{\rm m}(r) + \overline{\Sigma}_{\rm m}(r)\right)\right]^{1/2} \\
%	&  \propto & \left\{\frac{1}{r^3}\left[2M^\prime(r)+M(r)\right]\right\}^{1/2} \\
%	& \propto & \left[2\bar{\rho}^\prime(r)+\bar{\rho}(r)\right]^{1/2},
%\end{eqnarray*}
\begin{equation}\label{eqn:kappa_relation}
	\kappa(r) = \left[\frac{\pi G}{R}\left(2\Sigma_{\rm m}(R) + \overline{\Sigma}_{\rm m}(R)\right)\right]^{1/2}.
	%	\propto  \left[2\bar{\rho}^\prime(r)+\bar{\rho}(r)\right]^{1/2},
\end{equation}
%where $\bar{\rho}(r)$ 
%is the volume density by averaging the total mass $M(r)$ over 
%a sphere with a radius of $r$, and $\bar{\rho}^\prime$ is the volume density 
%calculated by using $\Sigma_{\rm m}(r)$ intead of $\overline{\Sigma}_{\rm m}(r)$ 
%to calculate the total mass within $r$.
This correlation indicates that the epicyclic frequency of the resolved region is positively correlated with the local mass surface density and the average surface density within the corresponding galactic-centric radius, while being negatively correlated with the radial distance from the center of the galaxy.
Similarly, one can also obtain the 
scaling relation for the Oort constant $A$ from its definition as
\begin{equation}\label{eqn:A_relation}
	A(R) = \frac{1}{2}\left[\frac{\pi G}{R\overline{\Sigma}_{\rm m}(R)}\right]^{1/2}
	\left[\frac{3}{2}\overline{\Sigma}_{\rm m}(R)-\Sigma_{\rm m}(R)\right].
\end{equation}
Similar to $\kappa$, $A$ also depends on both the local mass density and 
the average density within $R$, but in a more complicated way. 

\begin{figure*}
	\centering
	\includegraphics[width=\textwidth]{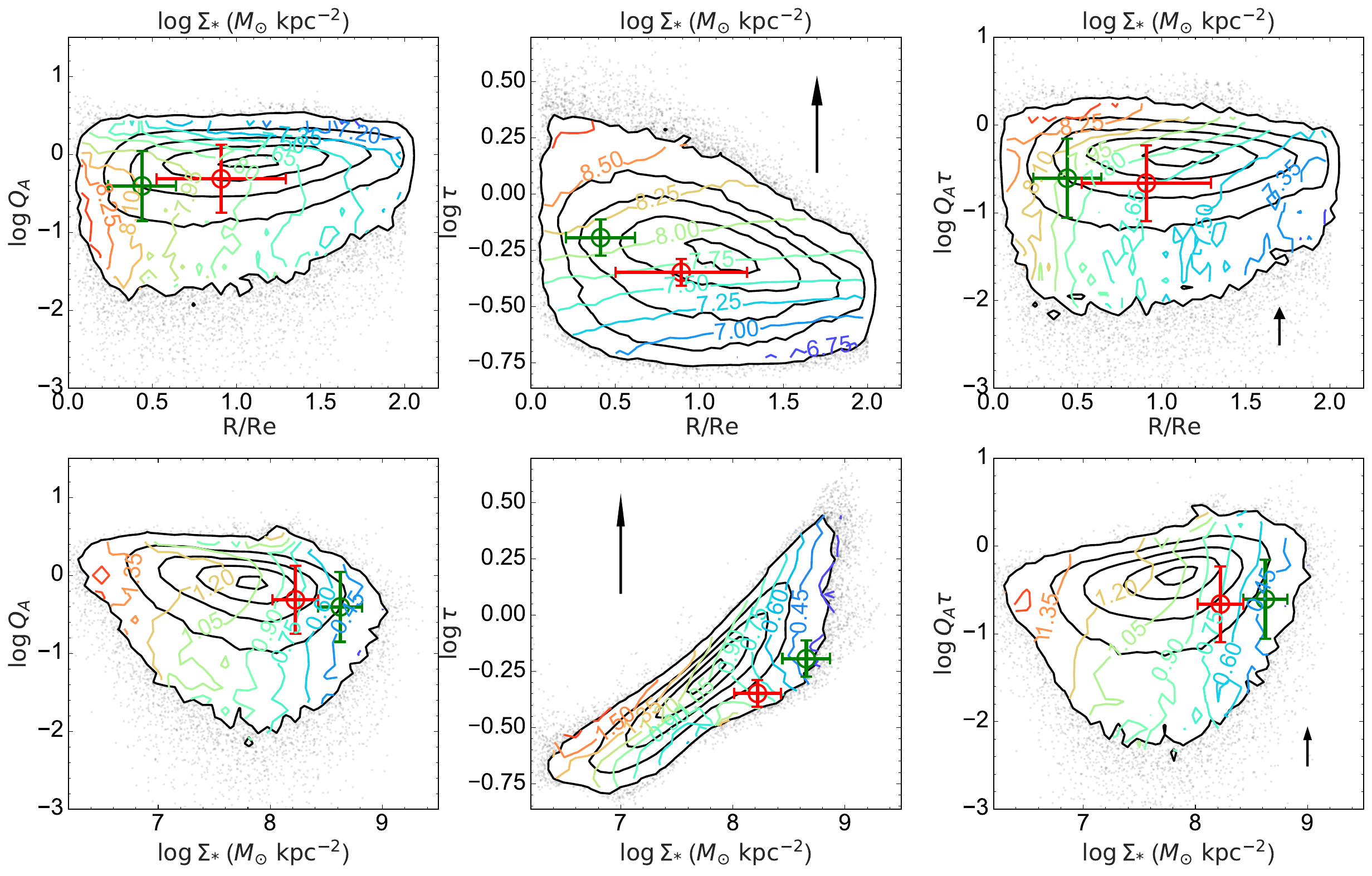}
	\caption{The upper panels (from left to right) display the radial profile of $\log Q_A$, $\log \tau$, and $\log Q_A \tau$, with the distribution of $\log \Sigma_\ast$ indicated by the colorful contours in each panel. The lower panels show the correlation of $\log Q_A$, $\log \tau$, and $\log Q_A \tau$ with $\log \Sigma_\ast$, with the distribution of $R/R_e$ shown as colorful contours in each panel. The black arrows indicate the systematic shift of $\log \tau$ and $\log Q_A \tau$ when the N2O2 method instead of the O3N2 method is utilized to estimate $Z_{\rm gas}$.}\label{fig:Qtau}
\end{figure*}

To better understand the role of \Sigstar, we have estimated 
$\kappa$ and $A$ for each spaxel in our sample. Details of the estimation 
process can be found in Appendix \ref{sec:app_dynamics}. In short, we firstly 
derive the rotation curve $v(R)$ for the host galaxies from the line-of-sight 
H$\alpha$ velocity maps, and then estimate both $\kappa$ and $A$ 
from $v(R)$ according to their definitions. In \autoref{fig:kappa}, we 
compare the estimates of $\kappa$ (left panel) and $A$ (center panel)
with the values given by the relations in \autoref{eqn:kappa_relation} 
and \autoref{eqn:A_relation}. For the x-axes of these two panels, we use $\Sigma_{\rm b}=\Sigma_{\rm gas}+\Sigma_\ast$ instead of $\Sigma_{\rm m}$, thus ignoring the 
contribution of dark matter which is not available from our data. 
Overall the distribution of all spaxels roughly follows the 1:1 relation
in both panels, with systematic offsets and large scatters 
which should be attributed to the omittance of dark matter and the 
oversimplified assumption of circular rotation in deducing the scaling 
relations, as well as the errors in the relevant measurements.
The distribution of $\log$ \Sigstar\ are shown as the colored contours.
We see a trend for both $\kappa$ and $A$ to increase with \Sigstar, 
indicating the positive correlation between existing stars and enhanced critical 
density. In the right panel of the same figure, we show the distribution 
of $Q_A$ and $Q_{\kappa}$, the $Q$ parameters defined above 
but calculated using $\kappa$ and $2.5A$ respectively. The two 
stability parameters show reasonably strong correlation with each 
other, although with large scatter. The contours of $\log$\Sigstar\ 
clearly show that the spaxels with highest \Sigstar\ are located in 
the ``subcritical'' regime defined by $Q_A<1$ and $Q_{\kappa}<1$.
This result suggests that, although a high density of stars leads to 
increased values for both $\kappa$ and $A$ as shown in the left 
two panels, the net effect of these stars is to decrease the $Q$ 
parameters, hence facilitating instability rather than suppressing it. 

In addition, in \autoref{fig:kappa} the median locations of GPQRs and 
GRQRs are plotted as the red and green circles, with the error bars 
indicating the 1$\sigma$ scatter of individual regions around the 
median. When compared to the parent sample, both types of quenched 
regions tend to have larger-than-average $\kappa$ and $A$, implying 
again the positive role of $\kappa$ and $A$ for quenching. 
In the $Q_A$ versus $Q_{\kappa}$ diagram the quenched regions 
are located slightly below the peak of the parent sample, with 
$Q_A\lesssim 1$ (slightly smaller than average) and $Q_{\kappa}\ga 1$
(just similar to average). This result implies that dynamical stabilization 
alone cannot fully drive the quenching of star formation in our galaxies, 
and that the high importance of \Sigstar\ is not pointing us to 
dynamical stabilization as the underlying mechanism for quenching. 

Based on standard disk instability analysis and zoom-in simulation 
technique, \citet{2009ApJ...707..250M} proposed the scenario of 
``morphological quenching'' that a prominent bulge, or more generally 
a centrally-concentrated mass distribution could stabilize the gas disk 
and quench star formation by sheared perturbations within the disk. 
Consequently, the galaxy could maintain weak or no star formation despite 
the presence of significant amounts of gas. In this case, the GRQRs 
studied in our work should be more commonly found in galaxies with 
a more concentrated morphology. However, we find the \sersic\ index, 
concentration, $\Sigma_{1kpc}$, or B/T of the host galaxies of GRQRs 
are all similar to those of the host galaxies of GPQRs (see \autoref{fig:compare_global}). 
% In addition, as seen from \autoref{fig:kappa}, the local shear rate $A$ is 
% primarily determined by the local stellar mass density $\Sigma_{\ast}$ 
% rather than the internal average density $\overline{\Sigma}_{\rm b}$, 
% implying that a centrally-concentrated morphology is not an important 
% factor. 
This is probably not unexpected because the concept of morphological 
quenching was initially proposed to understand the quenching in early-type 
galaxies, but not the spirals considered here.

\subsubsection{Self-shielding of gas}\label{sec:gas_role}

In addition to forming stars by dynamical instability as discussed above, 
the gas should be able to shield itself from background ionizing radiation 
so that it can cool down, condense and form stars \citep{Stecher1967, Federman1979, deJong1980, Glassgold1985, Franco1986, Draine1996, Bertoldi1996, Schaye2004, Krumholz2009,Vazquez-Semadeni2010, Krumholz2012, Sternberg2014, Ballesteros-Paredes2020, Sternberg2021, Maillard2021}.
In such models, the ability for gas to self-shield is determined by the dust 
optical depth, $\tau$, which can be estimated by \citep{2018MNRAS.474.2323S}
\begin{equation}
	\tau=\frac{\Sigma_{\rm gas}Z_{\rm gas}}{\left(\Sigma_{\rm gas}Z_{\rm gas}\right)_{ss}},
\end{equation}
where $Z_{\mathrm{gas}}$ represents the gas-phase metallicity relative to 
solar, and $\left(\Sigma_{\mathrm{gas}}Z_{\mathrm{gas}}\right)_{\mathrm{ss}}$ 
is the dust optical depth above which the gas becomes self-shielded. 
We estimate $\tau$ for each spaxel in our sample, using our measurements 
of \Siggas\ and 12+$\log$(O/H) (the O3N2-based gas-phase metallicity calibrated by \citealt{gas_z_O3N2})
and following \cite{2018MNRAS.474.2323S} to calculate the 
metallicity-dependent  $\left(\Sigma_{\mathrm{gas}}Z_{\mathrm{gas}}\right)_{\mathrm{ss}}$. We note that the O3N2 metallicity estimator is only reliable for regions primarily ionized by OB stars. It is appropriate for the majority of spaxels in our sample, providing a sufficient representation of the overall trend. Nevertheless, the derived metallicity of the quenched region may be unreliable. Although we present the measurements of the quenched region for comparison, it is important to understand the results with caution.
 
In \autoref{fig:Qtau} (center column) we show the estimates of $\tau$ against 
$R/R_e$ (first row) and \Sigstar\ (second row) for all the spaxels in the parent sample, with the distribution of \Sigstar\ (first row) and $R/R_e$ (second row) shown additionally as the colored contours. For comparison, the distribution 
of $Q_A$ against $R/R_e$ and \Sigstar\ is shown in the left column of the same figure. 
We use $Q_A$ rather than $Q_{\kappa}$ for this figure considering that 
the rotation curves of our galaxies are still rising over the radial range 
covered by the spaxels in our sample (see \autoref{fig:rotation_fit} for 
examples), but our conclusion would remain unchanged if $Q_{\kappa}$ 
was instead used. In the right column we show the product, $Q_A\tau$ 
for the same samples. As suggested by \citet{2018MNRAS.474.2323S}, 
the product of the $Q$ parameter and $\tau$ can provide useful constraints 
on whether fragmentation or opacity is the underlying driver of star formation. 
In the center and right panels for each row, the vertical arrow indicates the average 
amount that $\tau$ or $Q_A\tau$ would increase if the N2O2 method 
was instead adopted to estimate $Z_{\rm gas}$.  In all the panels 
the median location and the scatter of the two quenched samples are 
indicated by the red and green circles with error bars. 

The negative correlation of $Q_A$ with \Sigstar\ is seen again from the 
left column. For $\tau$ we see a positive correlation with \Sigstar, which
is expected from the known relation between gas-phase metallicity and 
stellar mass in star-forming galaxies 
\citep{Tremonti2004,Rosales-Ortega2012,Sanchez2013,Barrera-Ballesteros2016,Yao2022} and the molecular gas main sequence \citep{Lin2019}. This relation suggests that the role of a high density of stars is mainly 
to indirectly enhance the opacity in the disk and thus make the gas more self-shielded. 
% Both $\tau$ and the product $Q_A\tau$ fall below unity in the majority of 
% the spaxels in the parent sample. This appears to suggest that the star 
% formation in our galaxies is mainly driven by disk instability rather than 
% gas self-shielding. 
% On the other hand, one may also interpret this result as evidence for the unshielded gas ($\tau<1$) as the driver for quenching in most regions in our sample. 
At the same time, we find that, although located 
at relatively small radii, the quenched regions including both GPQRs and 
GRQRs have similar values in $Q_A$, $\tau$ and $Q_A\tau$ to the average 
of the parent sample. This result essentially rules out both dynamical 
stabilization and unshielded gas as the driving mechanisms for quenching.
Moreover, the high importance of \Sigstar\ should be telling different 
mechanisms to what have been discussed so far.

\subsubsection{Stellar feedback}

Existing stars can provide energy or kinematic feedback to suppress 
star formation through radiation pressure, stellar wind and supernova 
explosion. We first consider radiation pressure which has long been 
proposed as an efficient feedback mechanism to suppress star formation 
(e.g. \citealt{1983MNRAS.203.1011E, 2005ApJ...630..167T, 2013MNRAS.433.1970F, 2016ApJ...829..130R}). 
Previous studies have primarily focused on OB stars due to their high luminosity 
and ionizing photon production. Consequently, pressure regulation models
\citep{2011ApJ...731...41O, 2011ApJ...743...25K} have linked the strength of 
radiation pressure to SFR rather than the local stellar mass density. 
Considering that the quenched regions are dominated by old stellar populations,
here we consider the ability for the radiation pressure from evolved stars 
(such as red giants) to play some role in quenching. Following 
\citet{2016ApJ...829..130R}, we consider a simple ``gas shell'' model 
in which the central star or star cluster is surrounded by a spherical gas 
shell. In this simple case, the gas shell feels both self-gravity and stellar gravity 
which facilitate gas collapse and star formation, as well as radiation pressure 
from the central star/cluster which supports the gas shell against collapse. 
The competition between gravitational force and radiation pressure can be 
quantified by the Eddington ratio $\epsilon_{\rm Edd}$, defined as the ratio 
of the pressure as caused by radiation and gravity, as follows:
\begin{equation}
		\epsilon_{\rm Edd} = \frac{2 k_{\rm abs} \Psi \Sigma_{\ast}}{\pi G c \Sigma_{\rm gas}(2 \Sigma_{\ast} + \Sigma_{\rm gas})},
\end{equation}
where $k_{\rm abs}$ is the fraction of stellar luminosity absorbed by the gas 
shell and $\Psi$ is the mass-averaged luminosity-to-mass ratio. For a typical 
GRQR in our sample which has  
$\Sigma_{\rm gas} \sim 2.5 \times 10^7~\mathrm{M}_\odot \mathrm{kpc}^{-2}$ and 
$\Sigma_{\ast} \sim 3 \times 10^8~\mathrm{M}_\odot \mathrm{kpc}^{-2}$, 
%$\epsilon_{\rm Edd}$ is approximated to be
%	\begin{equation}
%		\epsilon_{\rm Edd} \approx  0.03 k_{\rm abs} \frac{\Phi}{\operatorname{erg} \mathrm{s}^{-1} \mathrm{g}^{-1}}.
%	\end{equation}
%Considering a red giant with a mass of $1  \mathrm{M}_\odot$ and 
%$\Phi\sim200 \operatorname{erg} \mathrm{s}^{-1} \mathrm{g}^{-1}$, we estimate
%that the Eddington ratio $\epsilon_{edd} \approx 6 k_{\rm abs}$. 
we estimate that 
\begin{equation}
	\epsilon_{\rm Edd} \approx  0.06 k_{\rm abs} \frac{\Psi}{L_\odot / M_\odot}.
\end{equation}
Considering a red giant with a mass of $1\mathrm{M}_\odot$ and 
$\Psi\sim100 L_\odot / M_\odot$, we estimate that the Eddington ratio 
$\epsilon_{edd} \approx 6 k_{\rm abs}$. For the gas shell to be stable 
against gravity, we expect $\epsilon_{\rm Edd}>1$, which requires 
$k_{\rm abs}\ga 0.16$, i.e. about $1/6$ or more of the luminosity radiated 
from the star/cluster is absorbed by the gas shell. 

It is not straightforward to accurately calculate $k_{\rm abs}$. On one hand, the 
cross-section of gas depends on a variety of factors including its ionization state, 
chemical composition, dynamical state, and metallicity. On the other hand, gas can 
indirectly feel the radiation pressure through the interaction with dust grains which 
gain momentum by absorbing photons. Here, we consider two extreme cases. 
In one case, gas feels radiation pressure 
only through Rayleigh scattering. Based on the interpolation formula provided by \cite{1962ApJ...136..690D}, we estimate that $k_{abs} \approx 10^{-7}$, indicating 
that the gas hardly feels any radiation pressure. In the other case, all absorbed or 
scattered photons transmit their momentum to the gas. Under this assumption 
and using the observed $E(B-V)$ we estimate the $k_{\rm abs}$ in V band 
for each spaxels in our sample. For GRQRs, we find a median value of $k_{\rm abs} \approx 0.8$, indicating that radiation feedback is highly efficient in preventing 
gas collapse. The reality in GRQRs is likely closer to this case than the first case,
as the interaction between gas and dust is ubiquitous, and gas can even freeze onto 
dust grains during the initial stages of gas collapse \citep{2020A&A...638A.105Z}. 
In fact, complete momentum transfer from dust to gas has been implemented 
in the FIRE simulation \citep{2018MNRAS.480..800H}. 

\begin{figure}
	\includegraphics[width=0.45\textwidth]{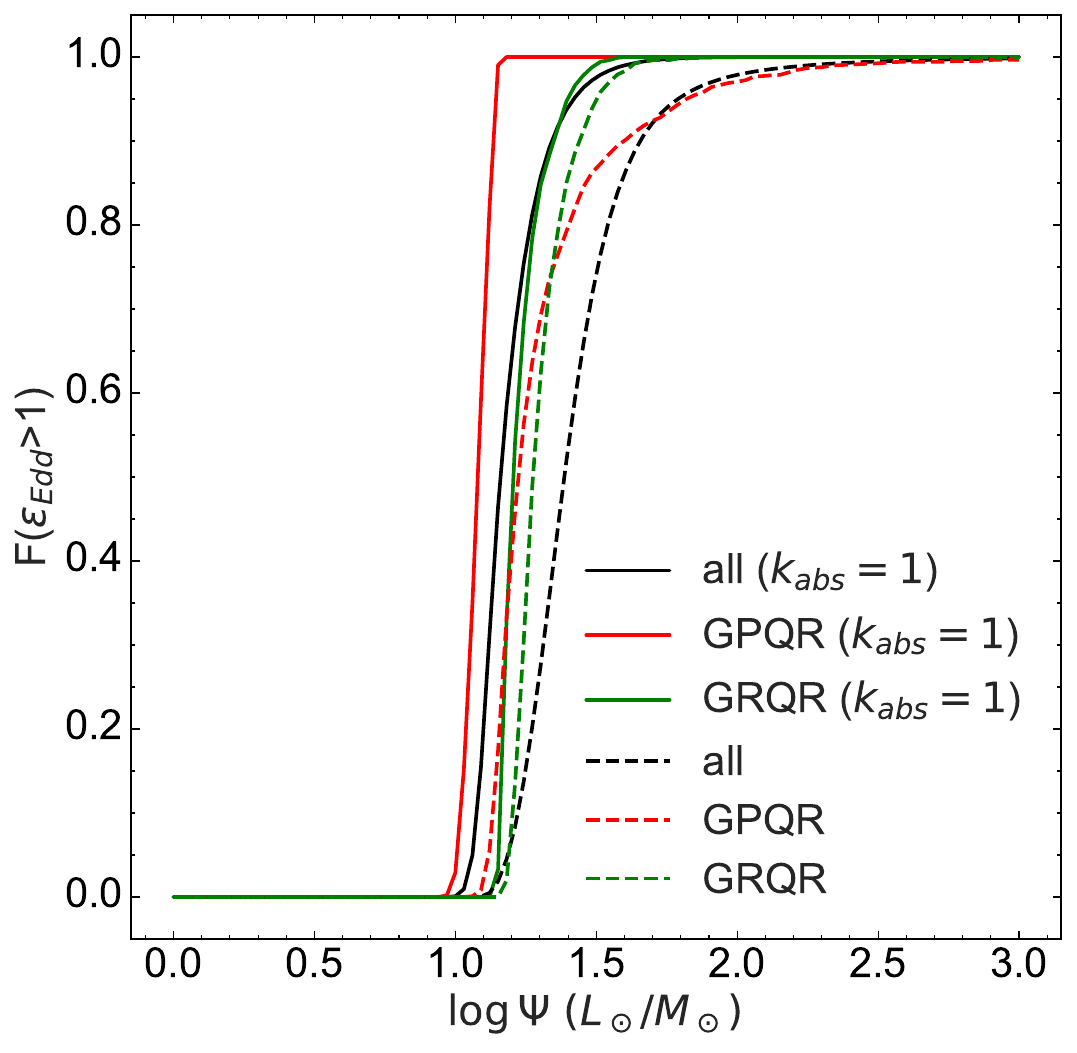}
	\caption{The $F(\epsilon_{\rm Edd}>1)$, the fraction of spaxels 
	with $\epsilon_{\rm Edd}>1$, as a function of $\Psi$ for the parent sample as 
	well as the two types of quenched regions. The solid lines show the results 
	obtained by assuming a fixed value of $k_{\rm abs}=1$, while the dashed lines show the results based on the estimates 
	of $k_{\rm abs}$ as derived from $E(B-V)$.}\label{fig:epsilon}
\end{figure}

\autoref{fig:epsilon} shows $F(\epsilon_{\rm Edd}>1)$, the fraction of spaxels 
with $\epsilon_{\rm Edd}>1$, as a function of $\Psi$ for the parent sample as 
well as the two types of quenched regions. The solid lines show the results 
obtained by assuming a fixed value of $k_{\rm abs}=1$. Since $k_{\rm abs}=1$
is the largest absorption fraction one can have, this case sets a lower limit 
of $\Psi$ that is required for $F(\epsilon_{\rm Edd}>1)$ to reach a given value
for a given sample. The dashed lines show the results based on the estimates 
of $k_{\rm abs}$ as derived above from $E(B-V)$.  For all the samples and 
in both cases of $k_{\rm abs}$, we see a sharp increase in $F(\epsilon_{\rm Edd}>1)$
when the luminosity-to-mass ratio of stars exceeds $\Psi\sim 10-20 L_\odot / M_\odot$. 
When the estimated $k_{\rm abs}$ are used, the $F(\epsilon_{\rm Edd}>1)$ 
of the GRQR sample reaches $\sim100\%$ at $\Psi\sim50 L_\odot / M_\odot$, which happens 
at $\Psi\ga 300 L_\odot / M_\odot$ for the other two samples. In other words, it seems that 
radiation pressure is more efficient, or more easy to take effect in GRQRs
than in GPQRs. The significant enhancement of the Gini importance of $\sigma_{\ast}$, a characteristic of aged stars, for GRQRs (green bar in \autoref{fig:RF_importance}) further supports the vital role of the old stellar population in GRQRs. This analysis confirms that radiation from the 
existing stars is able to substantially support the gas against collapse, 
as long as their stellar luminosity-to-mass ratios are larger than a few $\times10\ L_\odot / M_\odot$. 
Considering that the lifetime of a solar mass red giant is approximately 1 billion years, which is comparable to or even longer than the main sequence lifetime of more massive stars, red giants represent a plausible source of radiation pressure that is needed here to compete with self-gravity of the gas shell.  The presence of significant numbers of red giants naturally occurs in the quenched regions of our sample which are populated by high densities of old stars (see \autoref{fig:N2Ha_correlation}).
We strongly 
suggest the effect of radiation pressure from evolved stars, particularly red 
giants to be considered in future theoretical analyzes and hydrodynamical 
simulations of star formation.

Next, we consider stellar wind and supernova feedback. Stellar wind from 
massive OB stars and asymptotic giant branch (AGB) stars may eject dust 
and metal rich material to 
the ISM. This process releases energy and momentum, which can heat and 
expel surrounding gas.  Similar to stellar wind, supernovae (SNe) can not 
only inject energy and momentum to heat or expel the surrounding gas, 
but also result in the mixing of metal rich material and gas, thereby enhancing 
the cooling efficiency of the gas (\citealt{2019MNRAS.487.4393S}). 
Type Ia SNe exhibit a smooth distribution over time (\citealt{2019MNRAS.487.4393S}),
while type II SNe are associated with massive OB stars which should be less 
important in the quenched regions dominated by old populations. It is worth 
to note that both stellar wind and SN feedback can lead to higher N2H$\alpha$ 
and increased $\sigma_{\rm H\alpha}$, and the effect is expected to be proportion 
to $\Sigma_{\ast}$. Thus, their contributions can not be simply ruled out.
However, the wind from AGB stars is not highly efficient due to its low velocity, 
despite having a relatively high mass loss ratio. It becomes significant only 
when the adjacent gas exhibits a velocity dispersion greater than 
$300\ \mathrm{km s^{-1}}$ \citep{2015ApJ...803...77C}. Given the relative low 
velocity dispersion measured in our sample, with 
$\sigma_{\rm H\alpha}<100$ km s$^{-1}$ and $\sigma_\ast<200$ km s$^{-1}$ 
in most regions (see \autoref{fig:Q_tracer_correlation}), we can well expect 
the stellar wind to be inefficient in our case.

In order to estimate the effect of type Ia SN feedback in our sample, we utilize 
a model similar to that considered above for radiation pressure. The three 
forms of stellar feedback are similar in the sense that they transfer momentum 
and energy from the central star outwards, through photons (radiation pressure) 
or baryonic matter (stellar wind), or both (SN feedback). The effective Eddington 
ratio for type Ia SN feedback can be given by
\begin{equation}
	\epsilon_{\rm SN} = \frac{2 P_{\mathrm{SN}} \Sigma_{\ast}}{\pi G \Sigma_{\rm gas}(2 \Sigma_{\ast} + \Sigma_{\rm gas})},
\end{equation}
where $P_{\mathrm{SN}}$ represents the momentum injected into the ISM per 
unit time per unit stellar mass. By employing the Type Ia SN feedback model as 
implemented in \cite{2018MNRAS.480..800H} and assuming that all momentum 
released by Type Ia SNe can be coupled with the ISM, we obtain 
$P_{\mathrm{SN}} \sim 6.3 \times 10^{-4}\ \mathrm{km}/ \mathrm{s} /\mathrm{Myr}$. 
Considering the typical stellar and gas surface density of GRQRs, this yields
$\epsilon_{\rm SN} \approx 0.002$, suggesting a minor role of type Ia SN 
feedback in driving the quenching of GRQRs in our sample.

\subsubsection{Other mechanisms} \label{sec:other_mechanisms}

Previous studies have proposed some other quenching mechanisms, 
e.g. major mergers, violent disk instability and compaction quenching 
(efficient for gas-rich disk galaxies at high redshift), ram-pressure 
stripping and tidal stripping of gas (efficient for satellite galaxies orbiting 
in dark matter halos), active galactic nucleus feedback and virial shock 
heating (efficient for central galaxies in massive halos). As mentioned, 
we have removed interacting/merging galaxies from our sample, as well 
as those with misaligned H$\alpha$ distribution. In addition, the two 
types of quenched regions are shown to be hosted by non-AGN 
galaxies. Therefore, all the mechanisms listed above except gas 
stripping should not be important in our case. For gas stripping,
previous studies have shown that this effect depends strongly on 
the surface mass density of galaxies, with high-density galaxies to 
be more capable of sustaining their gas against stripping 
\citep[e.g.][]{Li2012,Zhang2013}. If gas stripping is at work for 
our galaxies, one would expect a negative role of \Sigstar\ for 
quenching, which is opposite to our finding. In addition, \citet{Li2023} 
have recently identified from SDSS a unique but 
rare population of globally quenched \hi-rich galaxies which 
are predominantly the central galaxies of relatively low-mass halos.
The quenching of these galaxies was attributed to the low surface 
density of the \hi\ disk as inferred from the extremely large 
\hi-to-optical disk size ratios, which could be supported by 
high angular momentum inherited from the high spin of host 
dark matter halos \citep[e.g.][]{Lutz2018}. This mechanism, advocated 
as ``angular momentum quenching'' by several authors 
\citep[e.g.][]{Obreschkow2016,Peng2020}, is unlikely a plausible
origin of the quenched regions in our sample which are mostly 
normal or even rich in terms of local gas density. 

\subsubsection{Summary of the subsection} 

To summarize, in this subsection we have discussed in depth the 
effect of dynamical stabilization, unshielded gas, and stellar feedback, 
which have been commonly considered in previous studies as 
plausible mechanisms for quenching. In particular, we have attempted 
to link the high density of stars as traced by \Sigstar\ with these mechanisms, 
trying to understand the role of existing evolved stars. We conclude 
that both dynamical stabilization and unshielded gas are unlikely 
to play an important role, while radiation pressure from evolved 
stars (likely dominated by red giants) can be efficient in providing
substantial support for the surrounding gas to be stable against 
gravitational collapse. 

\subsection{Limitations and outlook} \label{sec:limit}

This study is based on MaNGA data with an angular resolution of approximately $2^{\prime\prime}.5$, corresponding to $\sim$2 kpc at $z=0.04$, the redshift typical for MaNGA galaxies. This restricted resolution results in the point spread function (PSF) effect on the calculated $\Sigma_{1\mathrm{kpc}}$. For a test on this issue, we have attempted to calculate an image-based $\Sigma_{1\mathrm{kpc}}$ for each galaxy by applying the stellar mass-to-light ratio $M_\ast/L$ versus $g - r$ color relation as calibrated in \cite{Du2019} to the SDSS images which have a typical resolution of  $\sim1^{\prime\prime}.4$. We find consistent measurements for $z\lesssim0.02$, but at higher redshifts $\Sigma_{1\mathrm{kpc}}$ from MaNGA is underestimated with respect to the SDSS images due to limited resolution, as expected. We have repeated the randome forest analysis using the image-based $\Sigma_{1\mathrm{kpc}}$ measurements and found our results remain unchanged. This implies that the feature importance ranking is robust to the uncertainties in $\Sigma_{1\mathrm{kpc}}$. In addition, the limited resolution is also insufficient to resolve the star forming regions, thus the reported results should be regarded as the statistical average within approximately 2 kpc. The gas shell model assumed in our discussion is applicable for the scale of molecular clouds. The MaNGA resolution could blur the small-scale clumpiness, hence resulting in the underestimation of stellar/gas surface densities and gravitational potential. Conversely, the stellar/gas surface densities can also be overestimated, if multiple molecular clouds overlap along the same line of sight. It should be noted that not only the statistical results can vary with the resolution, as demonstrated by \cite{Pessa2022}, but different mechanisms can also operate at different spatial scales. Further investigation into the origins of the quenched region and the role of existing stars in star formation requires observations with spatial resolution comparable to that of molecular cloud scales.

This work mainly focuses on the regions within the disk component of isolated late-type galaxies. It is currently unclear whether the results reported in this paper are applicable to other components of late-type galaxies and all types of galaxies. Therefore, the future study should be extended to encompass all components in all types of galaxies in order to determine if the results are dependent on the morphology of the host galaxy. Additionally, it will allow a comprehensive comparison with the findings reported in existing literature (e.g., \citealt{Bluck2020-global-local,Bluck2020-central-satellite}).

The measurement of \Siggas\ is based on the estimator described in \autoref{sec:H2_measure}. Although the test in \autoref{fig:residual} shows that this estimator is unbiased on the key parameters included in the analysis, it still carries the potential risk of yielding artificial results. We have verified that the ranking of properties by random forest importance remains largely unchanged if we exclude \Siggas\ from the analysis. Additionally, we have attempted to reproduce the key plots (the right panel of \autoref{fig:GRQRs} and the second panel of \autoref{fig:Q_tracer_correlation_2d}) of this work in the EDGE-CALIFA sample, the sample used to calibrate the estimator. We find that both the existence of two types of quenched regions and the location of these regions in the \Siggas\ versus \Sigstar\ diagram are consistent with those reported in this work. However, it is imperative to use a much improved gas estimator or CO-derived molecular gas in future work in order to quantitatively verify or correct the results obtained in this study.

The findings of this study suggest that the high \Sigstar, as the most important property for quenching, could simultaneously contribute to the decrease of $f_{\mathrm{gas}}$ and SFE for quenched regions, especially for GRQRs. However, direct measurement of SFE in quenched regions is unattainable due to the contamination of ionization sources other than OB stars. Future research should employ a robust method to decompose the emission flux contributed by different sources, thus directly verifying the decrease of SFE with higher \Sigstar\ as proposed in this study. Additionally, such a decomposition method could also benefit the further study of different quenching definitions and their effects on the obtained results, as discussed in \autoref{sec:diff_definition}.

\section{Summary} \label{sec:Conc}

In this work, we use a sample of isolated disk galaxies selected from the 
MaNGA survey to investigate the significance of local/global properties 
of galaxies to the cessation of star formation at kpc scales.
We consider a total of 15 parameters, including both local properties 
for each of the resolved regions in MaNGA datacubes and the global 
properties for the host galaxies, as measured from the integral field 
spectroscopy data from MaNGA and multiwavelength observations from 
other surveys. We identify quenched regions by a single-parameter criterion
\Qtracer$ > 1.6-\log{2}=1.3$, thus requiring the quenched regions 
as having no/weak star formation not only at the present day but also 
during the past 1-2 Gyr. We then divide the quenched regions into two 
subsamples according to the surface mass density of cold gas \Siggas: 
gas-rich quenched regions (GRQRs) and gas-poor quenched regions (GPQRs). 
We first explore the global properties of the host galaxy of GRQRs and GPQRs. Subsequently, we utilize a random forest (RF) classifier to classify whether a resolved region 
is quenched or not given all the local/global properties, thus assessing 
the feature importance (FI) for each property. 
We then proceed to train RF regressors and classifiers to further evaluate the accuracy of different combinations of the properties in predicting the quenching parameter \Qtracer\ and quenching state.
Finally, we examine the dependence of our results on quenching 
definition by repeating the analysis of feature importance with quenched regions defined following \citet{Lin2019-quenching} and
\citet{Bluck2020-global-local,Bluck2020-central-satellite}.

Our conclusions can be summarized as follows.

\begin{enumerate}
	\item The quenched regions including both GRQRs and GPQRs tend to be 
	hosted by non-AGN galaxies with relatively high masses 
	(\mstar$\ga 10^{10}\rm M_\odot$) and red colors ($NUV-r\ga 3$), 
	as well as low SFR and high central density at fixed mass, but spanning 
	wide ranges in other parameters that are similar to the parent sample. 
	This implies that the conditions responsible for GRQRs are largely 
	independent on the global properties of host galaxies. 
	\item N2H$\alpha$ is identified to be the most significant single parameter 
	associated with quenching, regardless of how the quenched regions 
	are defined. The importance of N2H$\alpha$ can be understood 
	as a result of gas ionization by photons from substantially large 
	amounts of old stars. 
	\item \Sigstar\ is the most important property for quenching, especially for GRQRs, as long as a low \haew\ is required for identifying quenched regions.
	\item The difference in feature importance for quenching, as derived from our sample in comparison to those from the previous study by \citet{Bluck2020-global-local,Bluck2020-central-satellite}, are partly caused by the different quenching definitions, specifically the different requirements on whether low \haew\ is required for identifying a quenched region.
\end{enumerate}

Understanding the origin of the ionization conditions of regions excluded by the \haew\ cut is crucial for interpreting the different results obtained when requiring \haew\ or not. It is currently unclear whether these regions are genuinely quenched or if they consist of star-forming regions contaminated by AGN. Since the quenched regions selected with the \haew\ requirement represent a subset of those selected without this requirement, our findings suggest that AGN feedback alone cannot be regarded as the exclusive mechanism for quenching. This study underscores the importance of carefully treating regions with high \haew\ when selecting and studying quenched regions. However, further investigation is necessary to shed more light on this issue.

We have discussed in depth the implications of our results on the 
origin of gas-rich quenched regions in our sample, particularly 
the role of existing evolved stars as indicated by the high importance 
of \Sigstar. We consider three mechanisms: dynamical stabilization, 
unshielded gas, and stellar feedback, which have been commonly 
considered in previous studies. We conclude that both dynamical 
stabilization and unshielded gas are unlikely to play major roles
in quenching, and radiation pressure from evolved stars (likely 
dominated by red giants) can provide substantial support for the 
surrounding gas to be stable against gravitational collapse. Thus, the high \Sigstar\ could contribute to the decreasing of $f_{\mathrm{gas}}$ and SFE at the same time, especially for GRQRs.

As indicated in \autoref{sec:limit}, this study has certain limitations. These include the limited spatial resolution of MaNGA, the exclusive focus on the disk component of isolated late-type galaxies, the measurement of \Siggas\ by estimators, and the lack of decomposition of the emission line contributed by different ionization sources. The findings and proposed physical mechanisms in this study require further verification, revision, and development in future research.
% However, pointed out at the end of \autoref{sec:discussion}, 
% our work is based on the MaNGA data which has a spatial resolution 
% much larger than the sizes of individual star-forming regions and 
% molecular clouds. One should be cautious when interpreting the 
% results from our work as well as the previous studies based on similar 
% data. IFU surveys with even higher resolutions would be needed 
% in future for a full understanding of the star formation cessation 
% in galaxies. 

\section*{Acknowledgement}
%\begin{acknowledgments}

We are grateful to the anonymous referee for his/her comments which have helped us to improve this paper. This work is supported by the National Key R\&D Program of China
	(grant NO. 2022YFA1602902), and the National Natural Science 
	Foundation of China (grant Nos. 11821303, 11733002, 11973030, 
	11673015, 11733004, 11761131004, 11761141012). 

Funding for the Sloan Digital Sky Survey IV has been provided by the
Alfred P. Sloan Foundation, the U.S. Department of Energy Office of
Science, and the Participating Institutions. SDSS-IV acknowledges
support and resources from the Center for High-Performance Computing
at the University of Utah. The SDSS web site is www.sdss.org.

SDSS-IV is managed by the Astrophysical Research Consortium for the
Participating Institutions of the SDSS Collaboration including the
Brazilian Participation Group, the Carnegie Institution for Science,
Carnegie Mellon University, the Chilean Participation Group, the
French Participation Group, Harvard-Smithsonian Center for
Astrophysics, Instituto de Astrof\'isica de Canarias, The Johns
Hopkins University, Kavli Institute for the Physics and Mathematics of
the Universe (IPMU) / University of Tokyo, Lawrence Berkeley National
Laboratory, Leibniz Institut f\"ur Astrophysik Potsdam (AIP),
Max-Planck-Institut f\"ur Astronomie (MPIA Heidelberg),
Max-Planck-Institut f\"ur Astrophysik (MPA Garching),
Max-Planck-Institut f\"ur Extraterrestrische Physik (MPE), National
Astronomical Observatories of China, New Mexico State University, New
York University, University of Notre Dame, Observat\'ario Nacional /
MCTI, The Ohio State University, Pennsylvania State University,
Shanghai Astronomical Observatory, United Kingdom Participation Group,
Universidad Nacional Aut\'onoma de M\'exico, University of Arizona,
University of Colorado Boulder, University of Oxford, University of
Portsmouth, University of Utah, University of Virginia, University of
Washington, University of Wisconsin, Vanderbilt University, and Yale
University.
%\end{acknowledgments}

\appendix
\renewcommand\thefigure{\Alph{section}\arabic{figure}} % A1

\section{Measuring dynamical parameters} 
\label{sec:app_dynamics}

For the analysis in \autoref{sec:dynamics}, we have measured the 
epicyclic frequency $\kappa$ and the Oort constant $A$, from which 
we calculate the corresponding $Q$ parameters, $Q_{\kappa}$ and $Q_{A}$.
According to the definition of $\kappa$ (\autoref{eqn:kappa0}) and 
$A\equiv 0.5(v/R-dv/dR)$, both parameters include the local velocity 
gradient $dv/dR$. We obtain $dv/dR$ in two different ways. In the first 
method, for a given spaxel at coordinate $(R,\theta)$ where $R$ is the 
galactic-centric distance and $\theta$ is the position angle measured from the 
major axis of the host galaxy, we obtain its rotation velocity from the 
line-of-sight $\mathrm{H}\alpha$ velocity $v_{\mathrm{H\alpha, LOS}}$ 
by $v = v_{\mathrm{H\alpha, LOS}}/(\cos \theta \sin \mathrm{i})$, 
with $i$ being the inclination angle of the galactic disk given by 
the $r$-band minor-to-major axis ratio ($b/a$). The local velocity 
gradient is then estimated by 
\begin{equation}
	\frac{d v}{d R}=\frac{v_{\rm out}-v_{\rm in}}{\Delta R},
\end{equation}
where $v_{\rm in}$ is the rotation velocity of the spaxel in consideration,
and $v_{\rm out}$ is the rotation velocity of spaxels located 
at $(R+\Delta R,\theta)$. The radial interval $\Delta R$ is chosen 
appropriately, depending on the inclination angle, position angle, and the $R_e$-normalized 
spatial resolution of the host galaxy. Specifically, we create a set of mock line-of-sight velocity maps for galaxies with the same inclination angle, position angle, and $R_e$-normalized spatial resolution as the galaxies for which we will determine $\Delta R$. We employ a search algorithm to identify the $\Delta R$ that can reconstruct $\frac{dv}{dR}$ with the lowest mean-square-error, and subsequently apply this $\Delta R$ to real galaxies. Next, we estimate $\kappa$ and $A$ based on their definitions, utilizing the rotation velocity and local velocity gradient of each individual spaxel.

\begin{figure*}[ht!]
	\centering
	\includegraphics[width=0.9\textwidth]{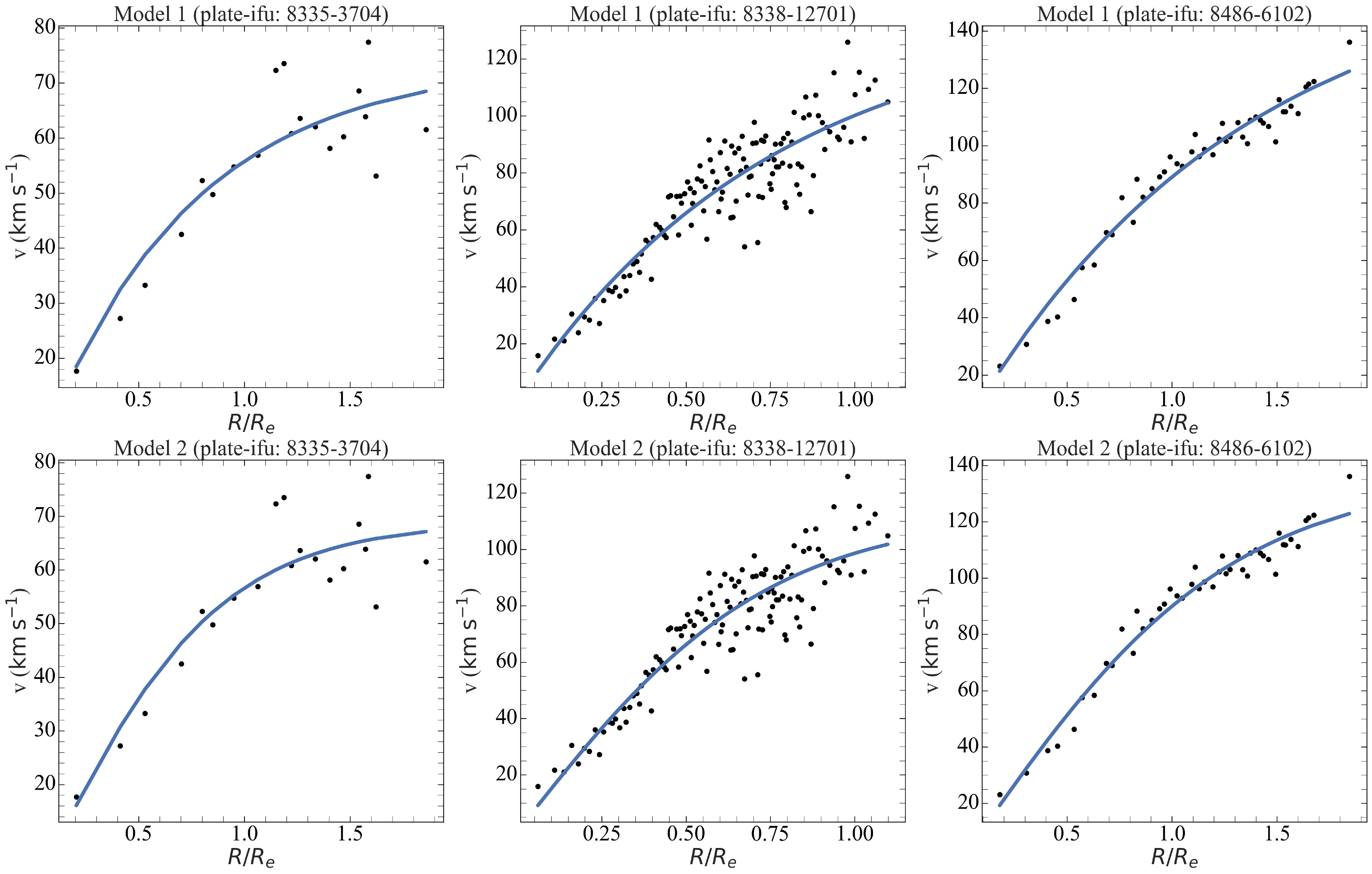}
	\caption{The examples of rotation curve fitting. Different columns represent randomly selected galaxies from our starting sample. The first row showcases the model $v(R) =v_{\mathrm{flat}}\left[1-e^{-R / l_{\mathrm{flat}}}\right]$, while the second row presents the model $v(R)=v_{\mathrm{flat}} \tanh \left(R / h_{\mathrm{rot}}\right)$. The black points in each panel denote every valid spaxel in the galaxy, with the blue curve representing the best-fitting result. \label{fig:rotation_fit}}
\end{figure*}

\begin{figure*}[ht!]
	\centering
	\includegraphics[width=0.8\textwidth]{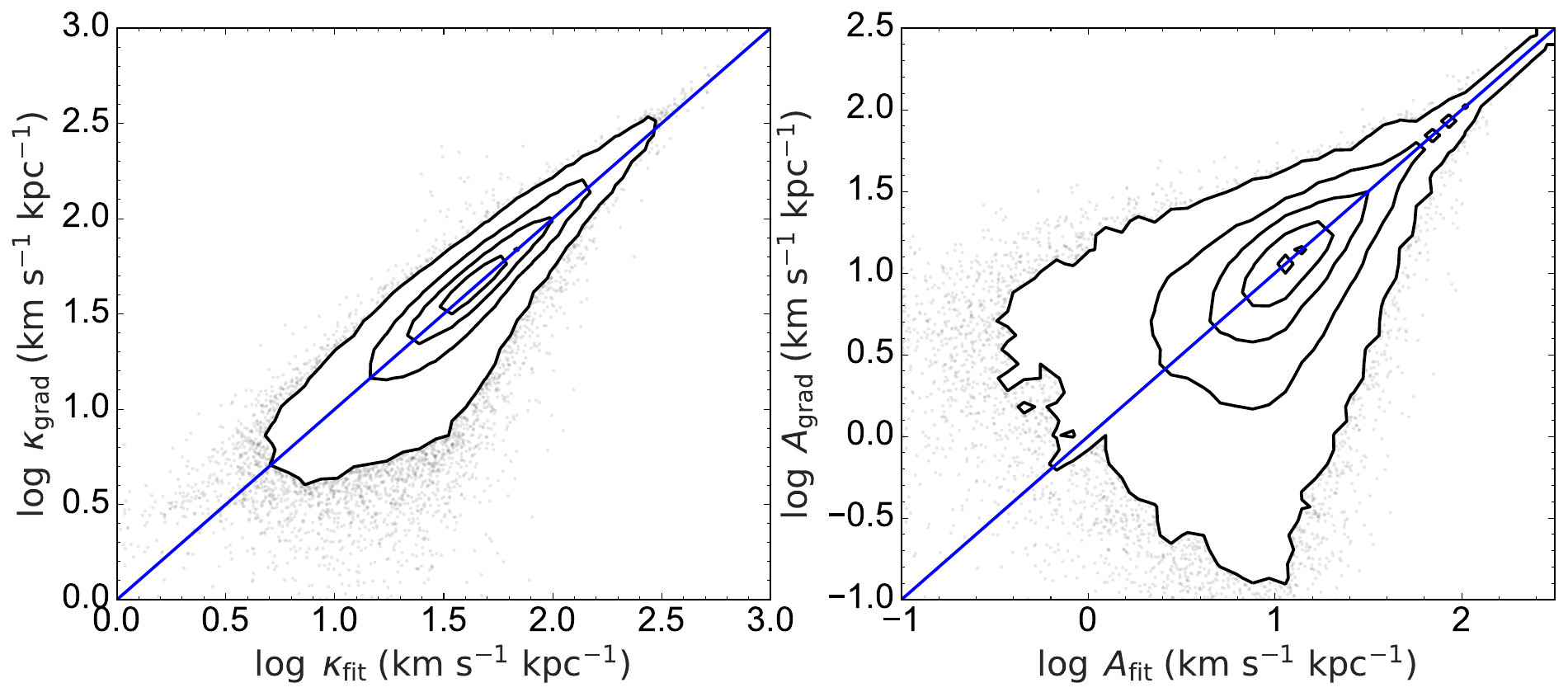}
	\caption{Consistency of $\kappa$ and $A$ obtained by rotation curve fitting and local gradient. The meanings of contours and black dots are same as \autoref{fig:GRQRs}. The blue diagonal line in each panel shows 1:1 correlation. \label{fig:twoQ}}
\end{figure*}

In the second method, we derive a rotation curve for the host galaxies, 
$v(R)$, defined as the average rotation velocity as a function of the 
galactic-centric distance $R$. For a host galaxy, we obtain $v(R)$ by fitting 
a functional form to the rotation velocities of all the spaxels with 
${\rm SNR}>3$ in the galaxy as function of their radial distances. 
Two rotation curve models used in \cite{2018MNRAS.474.2323S} 
are considered here: 
$v(R) =v_{\mathrm{flat}}\left[1-e^{-R / l_{\mathrm{flat}}}\right]$ and 
$ v(R)=v_{\mathrm{flat}} \tanh \left(R / h_{\mathrm{rot}}\right) $. 
We use both models to fit the data for a given galaxy and the one 
with a lower chi-square is adopted. Some examples of the rotation curve 
fits are shown in \autoref{fig:rotation_fit}. From the best-fit rotation 
curves we then calculate $\kappa$ and $A$ for each spaxel in our sample, 
following the definition of the two parameters. In 
\autoref{fig:twoQ} we compare the two methods of deriving $\kappa$ 
(left panel) and $A$ (right panel), which are in good agreement.
For the analysis in \autoref{sec:dynamics} we use the measurements 
from the first method, i.e. based on the local velocity gradients.

We then measure the modified \citet{1964ApJ...139.1217T} $Q$ parameter, $Q_{\kappa}$ following the definition by \citet[][also see \autoref{eqn:Q} 
and \autoref{eqn:Qgasstar}]{1994ApJ...427..759W}. For the characteristic 
sound speed $c_{\rm s}$ needed for calculating $Q_{\rm gas}$, we adopt 
a fixed value of $15~ {\rm km s^{-1}}$, which is the average velocity dispersion 
of ionized gas in the simulation of \citet{2009ApJ...707..250M}. We note that
a similar value of $13~ {\rm km s^{-1}}$ was adopted in 
\citet{2018MNRAS.474.2323S}. We follow \citet[][see their Eqn.6]{2018MNRAS.474.2323S} to indirectly estimate the radial stellar 
velocity dispersion, $\sigma_{\rm r,\ast}$, which is needed for calculating 
$Q_{\rm \ast}$. The surface stellar mass densities (\Sigstar) are measured 
from the MaNGA data, as described in \autoref{sec:DandM}. The surface 
gas mass densities (\Siggas) are estimated using the method described 
below. 

\begin{deluxetable*}{lccc}
	% \tablenum{1}
	\tablecaption{The molecular gas mass surface density fraction fitting results\label{tab:fit_res}}
	\tablewidth{0pt}
	\tablehead{
		\colhead{Variable} & \colhead{First Fit Coefficient} & \colhead{Second Fit Coefficient} & \colhead{Coefficient Converted Back to Original Variable} \\
	}
	\decimalcolnumbers
	\startdata
	const & -0.8697 & -0.8697 & 7.52670997\\
	$\log{\rm EW(H\alpha)}$ & -0.4576 & -0.4841 & -1.23631778\\
	$\log{\Sigma_{\ast}}$ & -0.6681 & -0.6885 & -1.44821091\\
	$\mathrm{E(B-V)}$ & -0.1276 & -0.0905 & -0.28691643\\
	$\log{\rm N2H\alpha}$ & 0.1271 & 0.103 & 0.69222\\
	$[\log{\rm EW(H\alpha)}] ^ 2$ & 0.0118 & - & -\\
	$\log{\rm EW(H\alpha)} \cdot \log{\Sigma_{\ast}}$ & 0.5035 & 0.5448 & 0.18309026\\
	$\log{\rm EW(H\alpha)} \cdot \mathrm{E(B-V)}$ & 0.0121 & - & - \\
	$\log{\rm EW(H\alpha)} \cdot \log{\rm N2H\alpha}$ & 0.0021 & - & -\\
	$(\log{\Sigma_{\ast}}) ^ 2$ & 0.3278 & 0.3509 & 0.04597756\\
	$\log{\Sigma_{\ast}} \cdot \mathrm{E(B-V)}$ & 0.1493 & 0.1029 & 0.03949803\\
	$\log{\Sigma_{\ast}} \cdot \log{\rm N2H\alpha}$ & -0.1086 & -0.0817 & -0.07043908\\
	$[\mathrm{E(B-V)}] ^ 2$ & -0.0182 & - & -\\
	$\mathrm{E(B-V)} \cdot \log{\rm N2H\alpha}$ & 0.0025 & - & -\\
	$[\log{\rm N2H\alpha}] ^ 2$ & -0.0017 & - & -\\
	\enddata
	%	\tablecomments{}
\end{deluxetable*}

\section{Estimating gas surface densities} \label{sec:H2_measure}

For the lack of spatial resolved CO and \hi\ observations for our sample, 
we indirectly estimate the molecular and atomic gas surface densities, 
\Sigmol\ and \SigHI, and thus obtain the total gas density by 
\Siggas$=$\Sigmol$+$\SigHI.

\begin{figure*}
	\plotone{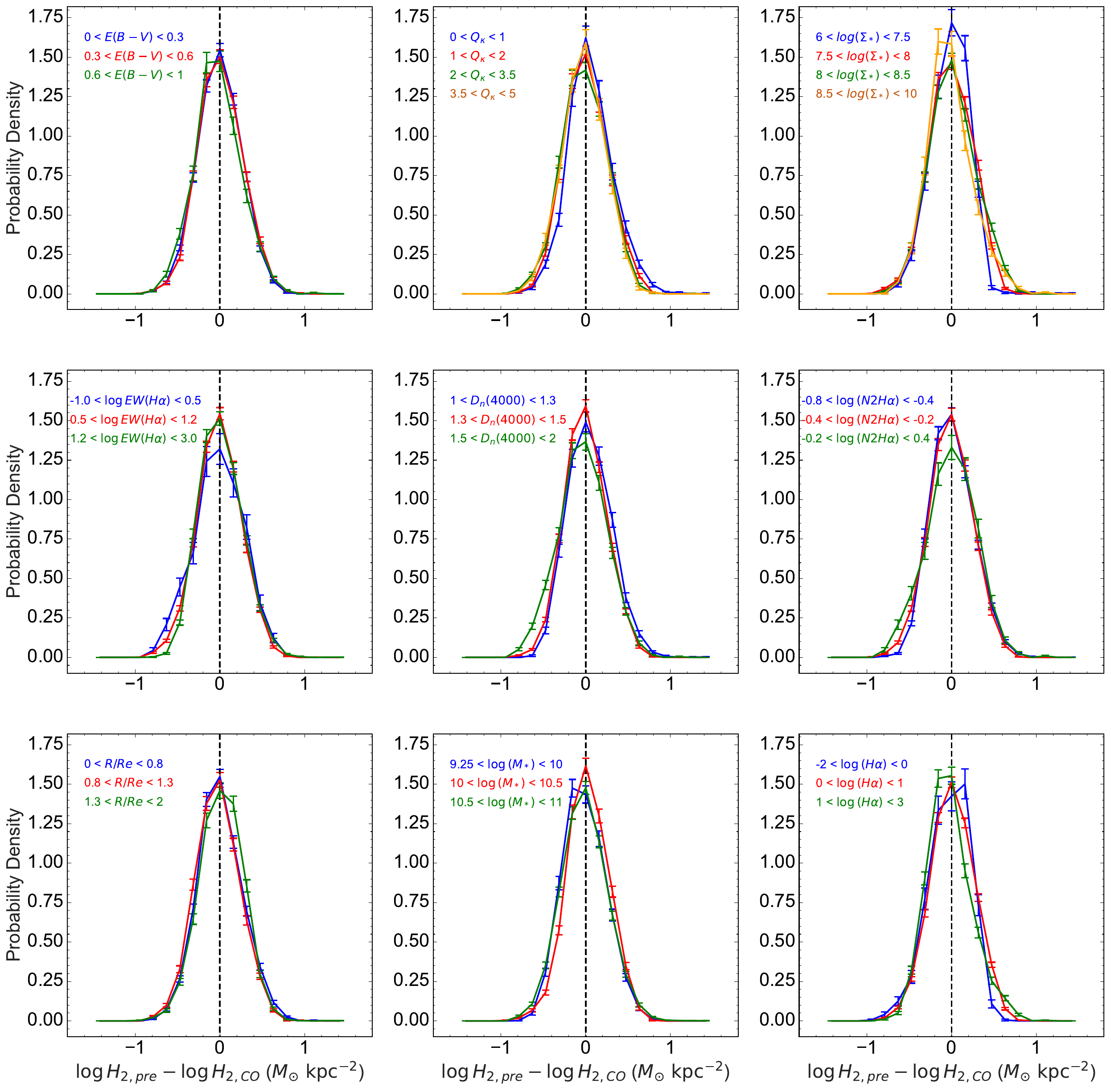}
	\caption{No bias of molecular gas estimator on parameters used in analysis. Different panels show the probability density distribution of residual ($\log{\Sigma_{H2_{ pre}}} - \log{\Sigma_{H2_{CO}}}$) in different bins (represented by different colors) of different parameters. The parameters shown in first row are E(B-V), $Q_{\kappa}$ and $\Sigma_{*}$. The parameters in second row are \haew, \dnbreak\ and N2H$\alpha$. The parameters in third row are radius, galaxy mass and corrected H$\alpha$ flux. The vertical line in each panel locates at zero residual as a reference. The error bar shows 1 $\sigma$ uncertainty. The $\sigma_{\rm r,\ast}$ used to calculate $Q_{\kappa}$ is approximated by line-of-sight $\sigma_{\ast}$ in this figure. All the test data used here are obtained from EDGE-CALIFA sample. \label{fig:residual}}
\end{figure*}

To estimate \Sigmol, we have developed an estimator for 
$f_{\rm H_2} \equiv \Sigma_{\rm H_2}/(\Sigma_{\ast}+\Sigma_{\rm H_2})$,
based on four parameters that we have measured from the MaNGA data
for our sample. These include the [N{\sc ii}]-to-H$\alpha$ flux ratio
(N2H$\alpha$), the nebular dust attenuation \ebv, 
the surface stellar mass density \Sigstar, and the equivalent width 
of the H$\alpha$ emission line \haew. We obtain and test the best-fit model 
of the estimator using data from the EDGE-CALIFA Survey 
\citep{2012A&A...538A...8S, 2014A&A...569A...1W, 2016A&A...594A..36S, Colombo2018}, which provides both MaNGA-like IFU data 
and resolved CO observations. We select a sample of spaxels following 
the selection of our MaNGA sample, and for each spaxel we preform 
spectral fitting in the same way as done for the MaNGA spectra
(see \autoref{sec:measurement}). Additionally, we require 
$\mathrm{\Sigma_{H_2}}$ to fall in the sensitive range of the EDGE survey, 
$4 - 110 \times 10^6 \ \mathrm{M_\odot /\ kpc^2}$ 
\citep{2020MNRAS.492.2651B}. The molecular gas mass is converted 
from the CO surface brightness, assuming a constant conversion factor 
$\alpha_{CO} = 4.4\ \mathrm{M}_{\odot}\left(\mathrm{K}\ \mathrm{km} \mathrm{s}^{-1} \mathrm{pc}^{2}\right)^{-1}$ \citep{2020MNRAS.492.2651B}. 
To obtain an estimator as accurate as possible, we take $\log({\rm N2H\alpha})$,
\ebv, $\log$\Sigstar, and $\log$\haew, as well as their pairwise quadratic 
combinations as independent variables. Considering the different numerical 
ranges and dimensions, all the variables are standardized before fitting. 
We then fit the measurements of $f_{\rm H_2}$ from the EDGE-CALIFA 
sample by a linear combination of all the independent variables. We remove 
the variables with the best-fit coefficient less than 0.1 and perform the 
fitting again using the remaining variables. The best-fit coefficients as obtained 
from the standardized variables are listed in \autoref{tab:fit_res}, for 
the two fitting runs. We use the best-fit relation from the second run 
as our estimator, for which the coefficients for the original variables are 
also listed in \autoref{tab:fit_res}. \autoref{fig:residual} shows the residual 
of the predicted $f_{\rm H_2}$ by our estimator relative to the measurement
from the EDGE-CALIFA survey. The different panels compare the results 
for subsamples divided by different properties as indicated. As can be 
seen, in all cases the residual distributions are centered at zero with a 
scatter of about 0.26 dex, independent on the properties considered. This independence ensures our estimator will not lead to artificial results at least in first order. This estimator is applied to the MaNGA data to estimate \Sigmol\ for each of the spaxels in our sample.

\begin{figure*}
	\centering
	\includegraphics[width=0.9\textwidth]{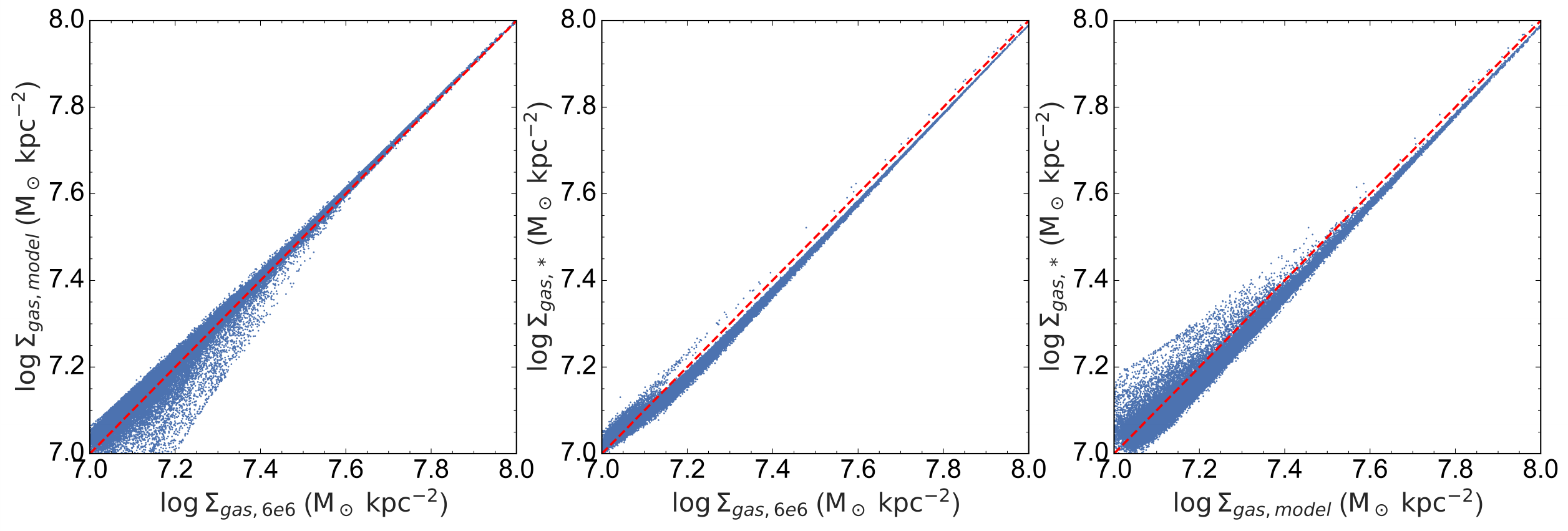}
	\caption{Consistency of total gas mass surface density in our sample by using different HI estimators. The three panels from left to right show the pairwise comparison of the following three estimators: the fixed value at $6 \times 10^6 M_\odot / \text{kpc}^2$ ($\Sigma_{gas, 6e6}$), \cite{2016MNRAS.460.2143W}'s model ($\Sigma_{gas, model}$) and stellar mass surface density-HI mass surface density relationship found by \cite{2020MNRAS.496.4606M} ($\Sigma_{gas, stellar}$). The blue dots in three panels are spaxels on which we use different HI estimators to obtain their total gas mass surface density. The diagonal red dash lines in these panels show 1:1 correlation. \label{fig:HI-compare}}
\end{figure*}

To estimate \SigHI, we consider three different methods. Firstly,
we estimate a radial profile of \hi\ surface mass for each galaxy in our sample,
using the \hi\ size-mass relation and average \hi\ surface mass profiles provided 
by \cite{2016MNRAS.460.2143W, 2020ApJ...890...63W}. For this we use 
the total \hi\ mass of our galaxies available from the HI-MaNGA data release 2 \citep{2019AAS...23336323M}. Secondly, we estimate the \hi\ 
mass density using the average \Sigstar-\SigHI\ relation found by 
\cite{2020MNRAS.496.4606M}. Thirdly, for all the spaxels we simply assume 
a constant value of $\Sigma_{HI} = 6 \times 10^6 \ \mathrm{M_\odot kpc^{-2}}$  \citep{2020MNRAS.492.2651B}. We have tried with all the three methods,
finding no significant differences in the estimated total gas mass density 
$\Sigma_{\mathrm{gas}}$ measurement. This finding is shown in \autoref{fig:HI-compare}. We notice that some spaxels substantially 
deviate from the $1:1$ relation, and we find the majority of these spaxels 
are hosted by low mass galaxies with $M_\ast< 10^{10} M_\odot$). 
This is not an important issue for our work. 
As shown in \autoref{fig:compare_global}, the quenched regions in our 
sample are mostly hosted by high mass galaxies with $M_\ast\ga 10^{10}M_\odot$.
Therefore, for simplicity we set \SigHI\ as a constant of 
$6 \times 10^6~ {\rm M_\odot kpc^{-2}}$ throughout this work.

\bibliography{merged}{}
\bibliographystyle{aasjournal}

\end{document}